\documentclass[twocolumn,tighten]{aastex63}%,linenumbers]{aastex63}
\usepackage{amsmath,amssymb}
\usepackage{color}
\usepackage{longtable}
\usepackage[T1]{fontenc}
\usepackage{ulem}
\usepackage{wrapfig}
\usepackage{appendix}

\usepackage{savesym}
\savesymbol{tablenum}
\usepackage{siunitx}
\restoresymbol{SIX}{tablenum}

\usepackage{booktabs,multirow}

\newcommand{\response}[1]{{#1}}%{\bf #1}}

\graphicspath{{./}{figures/}}

\begin{document}

\title{Matching Globular Cluster Models to Observations}

%% Note that the corresponding author command and emails has to come
%% before everything else. Also place all the emails in the \email
%% command instead of using multiple \email calls.
\correspondingauthor{Nicholas Z. Rui}
\email{nrui@caltech.edu}

\author[0000-0002-1884-3992]{Nicholas Z. Rui}
\affil{TAPIR, California Institute of Technology, Pasadena, CA 91125, USA}

\author[0000-0002-4086-3180]{Kyle Kremer}
\affiliation{TAPIR, California Institute of Technology, Pasadena, CA 91125, USA}
\affiliation{The Observatories of the Carnegie Institution for Science, Pasadena, CA 91101, USA}
\affil{Center for Interdisciplinary Exploration \& Research in Astrophysics (CIERA) and Department of Physics \& Astronomy, Northwestern University, Evanston, IL 60208, USA}

\author[0000-0002-9660-9085]{Newlin C. Weatherford}
\affil{Center for Interdisciplinary Exploration \& Research in Astrophysics (CIERA) and Department of Physics \& Astronomy, Northwestern University, Evanston, IL 60208, USA}

\author[0000-0002-3680-2684]{Sourav Chatterjee}
\affil{Tata Institute of Fundamental Research, Homi Bhabha Road, Navy Nagar, Colaba, Mumbai 400005, India}

\author[0000-0002-7132-418X]{Frederic A. Rasio}
\affil{Center for Interdisciplinary Exploration \& Research in Astrophysics (CIERA) and Department of Physics \& Astronomy, Northwestern University, Evanston, IL 60208, USA}

\author[0000-0003-4175-8881]{Carl L. Rodriguez}
\affil{Department of Physics, Carnegie Mellon University, Pittsburgh, PA 15213}

\author[0000-0001-9582-881X]{Claire S. Ye}
\affil{Center for Interdisciplinary Exploration \& Research in Astrophysics (CIERA) and Department of Physics \& Astronomy, Northwestern University, Evanston, IL 60208, USA}

\begin{abstract}
As ancient, gravitationally bound stellar populations, globular clusters are abundant, vibrant laboratories characterized by high frequencies of dynamical interactions coupled to complex stellar evolution.
Using surface brightness and velocity dispersion profiles from the literature, we fit $59$ Milky Way globular clusters to dynamical models from the \texttt{CMC Cluster Catalog}.
Without doing any interpolation\response{, and without any directed effort to fit any particular cluster}, $26$ globular clusters are well-matched by at least one of our models.
We discuss in particular the core-collapsed clusters NGC 6293, NGC 6397, NGC 6681, and NGC 6624, and the non-core-collapsed clusters NGC 288, NGC 4372, and NGC 5897.
As NGC 6624 lacks well-fitting snapshots on the main \texttt{CMC Cluster Catalog}, we run six additional models in order to refine the fit.
We calculate metrics for mass segregation, explore the production of compact object sources such as millisecond pulsars, cataclysmic variables, low-mass X-ray binaries, and stellar-mass black holes, finding reasonable agreement with observations.
Additionally, closely mimicking observational cuts, we extract the binary fraction from our models, finding good agreement except in the dense core regions of core-collapsed clusters.
Accompanying this paper are a number of \textsf{python} methods for examining the publicly accessible \texttt{CMC Cluster Catalog}, as well as any other models generated using \texttt{CMC}.
\end{abstract}

%% The \author command can take an optional ORCID.

\defcitealias{kremer2019cmcgrid}{K20}

\renewcommand{\vec}[1]{\mathbf{#1}}


%% Start the main body of the article. If no sections in the 
%% research note leave the \section call blank to make the title.

\section{Introduction}

Some of the oldest known structures, globular clusters (GCs) are gravitationally bound stellar populations located in galactic halos which formed $\sim13$ Gyr ago in the early universe \citep{hut2003gravitational}.
The pervasiveness and rich dynamical activity within globular clusters make them excellent sandboxes in which to study an abundance of stellar exotica, including X-ray binaries, radio millisecond pulsars, and gravitational wave sources
\citep{hui2010dynamical,ivanova2010formation,bae2014compact,kremer2018xrb,kremer2018lisa,ye2019millisecond}.

GCs are large $(N\gtrsim10^5$--$10^6)$ self-gravitating systems of objects with a large range of masses, for which dynamics is both complicated and critical to the formation and subsequent evolution of the cluster.
In recent years, some authors \citep{hut1992binaries,chatterjee2013understanding,kremer2020iau} have demonstrated that GCs which exhaust their supply of black holes undergo a runaway in core stellar density (``core collapse'') which is only stabilized by dynamical interactions between binaries (``binary burning'').
Core collapse is characterized observationally by a highly compact, bright core with a surface brightness profile which appears to constantly increase towards the GC's center, whereas the lack of core collapse is associated to a GC core with roughly flat surface brightness.
Today, roughly one-fifth of observed GCs in the Milky Way display the extreme central concentration in surface brightness characteristic of core collapse \citep[2010 edition]{harris1996catalog}.

%After violently relaxing over only a few crossing times \citep{lynden1967statistical}, GCs virialize into a state of quasi-equilibrium maintained by the ``burning'' of dynamical heat sources, most notably black holes and binaries \citep{heggie2003gravitational,kremer2020iau}.
%The exhaustion of such sources heralds a runaway in the core stellar density (``core collapse'') which is only stabilized when new dynamical heat sources are generated and segregated through stellar evolution and dynamical processes \citep{sigurdsson1993binary,chatterjee2013understanding,kremer2018black}.
%The concept of core collapse dates back to the pioneering work of \citet{antonov1962most} and \citet{lynden1968gravo} who showed that the central density of a self-gravitating large-$N$-body system formally diverges in finite time due to unmitigated outward energy flow caused by the system's negative dynamical heat capacity (in a so-called ``gravothermal catastrophe'').

Though in principle the most trustworthy method for GC dynamical modeling, direct $N$-body integration is extremely computationally expensive \citep[requiring, e.g., a year of supercomputing time for $N\simeq10^6$ particles,][]{wang2016dragon}, restricting its widescale application to star clusters with $N\lesssim10^4-10^5$  \citep[e.g.,][]{zonoozi2011direct,zonoozi2014direct} or requiring approximate ad hoc scalings with $N$ to realistic GC sizes \citep{aarseth1998basic, baumgardt2001scaling}.

Fortuitously, the introduction of more efficient methods, such as the Monte Carlo algorithm first introduced by H\'enon \citep{henon1971monte,stodolkiewicz1986dynamical,giersz1998monte,joshi2000monte}, has made possible the simulation of comprehensive GC model grids on realistic time frames \citep{rodriguez2016million}.
This development has kicked off extensive recent work on GC dynamics \citep[e.g.,][]{joshi2001monte,fregeau2003monte,fregeau2007monte,chatterjee2010monte,giersz2011monte,umbreit2012monte,giersz2013mocca,giersz2015mocca}, and singularly enables the analysis presented in this work.
In this work, we examine GC models generated by \texttt{Cluster Monte Carlo} (\texttt{CMC}), a H\'enon-style Monte Carlo code that computes the evolution of GCs under the assumption of spherical symmetry \citep{pattabiraman2013parallel}.

%To complement the upcoming release of \texttt{Cluster Monte Carlo} \citep[\texttt{CMC},][]{rodriguez2019cmc}, 
In particular, we explore the most recent grid of Milky Way GC dynamical models, the \texttt{CMC Cluster Catalog} \citep{kremer2019cmcgrid}, and present a procedure for determining a modern-day GC's location on the grid via its observed surface brightness and velocity dispersion profiles (SBPs and VDPs).
We summarize the \texttt{CMC Cluster Catalog} in Section \ref{modelgrid} and the fitting procedure in Section \ref{extractingclusterobservables}.
For concreteness, we specifically examine six of the GCs well-fit by this procedure, namely the core-collapsed clusters NGC 6293, 6397, and 6681 (Sections \ref{ngc6293}, \ref{ngc6397}, and \ref{ngc6681}), and the non-core-collapsed clusters NGC 288, 4372, and 5897 (Sections \ref{ngc288}, \ref{ngc4372}, and \ref{ngc5897}), which are all well-fit by the \texttt{CMC Cluster Catalog} as is.
We also consider NGC 6624, an interesting, high-metallicity cluster which is not captured initially on the \texttt{CMC Cluster Catalog} proper, and extend the model grid with additional \texttt{CMC} models to obtain a good fit (Section \ref{ngc6624}).
For these clusters, we consider exotic binary and millisecond pulsar populations, cluster masses and mass-to-light ratios (Section \ref{mml}), binary fractions (Section \ref{binary}), mass segregation (Section \ref{massseg}), and black holes (Section \ref{bh}).
Accompanying this work is a set of publicly available \textsf{python} functions and simulation properties needed to reproduce this analysis\footnote{https://github.com/NicholasRui/cmctoolkit} \response{\citep{codetag}}.

\section{Methods}

Here, we outline the methods used to compare our cluster models to the observed data of Milky Way GCs.
In Section \ref{modelgrid}, we broadly summarize the \texttt{CMC Cluster Catalog}, and, in Section \ref{extractingclusterobservables}, we describe the procedure for extracting the simulated surface brightness and velocity dispersion and using these to fit the observed data.
In Section \ref{exotica}, we detail criteria for identifying various stellar exotica in our models.

\subsection{Model Grid} \label{modelgrid}

The \texttt{CMC Cluster Catalog} comprises $148$ models spanning a realistic and comprehensive range of initial virial radii, tidal radii, metallicities, and masses (Table \ref{tab:params}), integrated via \texttt{CMC}.
Within \texttt{CMC}, stellar evolution is modeled using the \textsc{single-star evolution} \citep[\textsc{sse},][]{hurley2000comprehensive} and \textsc{binary-star evolution} \citep[\textsc{bse},][]{hurley2002evolution} algorithms updated to use the most current prescriptions for compact object formation \citep[see, e.g.,][]{Breivik2020}\response{.
These prescriptions describe the evolution of stars/stellar objects through various evolutionary stages which are distinguished in the code by ``startype'' (see Section 1 of \citealt{hurley2000comprehensive} for a list of startypes and discussion).
\texttt{CMC} also incorporates the physics of} three/four-body encounters are integrated using the \textsc{fewbody} package \citep{fregeau2004stellar,fregeau2007monte}, updated to include post-Newtonian terms \citep{rodriguez2018post}.
\response{Our models assume that GCs experience a constant tidal field throughout their lifetimes.
Of course, in general, GCs undergo complicated orbits characterized by periapse passages which may affect their dynamics in a nonlinear fashion, and there is evidence to believe that the galactic potential has itself varied over time in the history of the Milky Way \citep[e.g.,][]{kruijssen2019formation}.
Further exploration of the effects of such time-dependent tidal fields is beyond the scope of this work.}

\begin{table}
    \centering
    \begin{tabular}{l l}
    \hline
    Parameter & Values \\
    \hline
    Initial number of stars $N$ ($\times10^5$) & $2$, $4$, $8$, $16$, $32^*$ \\
    Virial radius $r_v$ (pc) & $0.5$, $1.0$, $2.0$, $4.0$ \\
    Galactocentric distance $R_g$ (kpc) & $2$, $8$, $20$ \\
    Metallicity $[\mathrm{M}/\mathrm{H}]$ & $-2$, $-1$, $0$ \\
    \hline
    \end{tabular}
    \caption{A summary of the initial cluster parameters of the \texttt{CMC Cluster Catalog}.\\ $^*$Due to computational expense, the grid only covers a subset of the allowed $r_v$, $R_g$, and $[\mathrm{M}/\mathrm{H}]$ values for $N=3.2\times10^6$.}
    \label{tab:params}
\end{table}

At various snapshots in time separated by multiples of the estimated dynamical time of the cluster, \texttt{CMC} writes a catalog of stellar and kinematic properties for all stars in the cluster.
In this study, we are primarily interested in cluster models similar to the old GCs observed in the Milky Way, thus we restrict our attention only to snapshots for which $t>10$ Gyr (all models are run to $t\approx14$ Gyr or until dissolution), of which there are 7,537 throughout the entire \texttt{CMC Cluster Catalog}.

In general, the parameters most germane to dynamical structure are the initial number of stars $N$ and the virial radius $r_v$.
The galactocentric distance $R_g$ is most impactful through its influence at the outskirts of the cluster as it defines the tidal radius, while the metallicity $Z$ primarily influences stellar evolution.
Moreover, both $R_g$ and $Z$ are more easily estimated empirically than $N$ and $r_v$, which often change drastically over the course of a cluster's lifetime \citep[e.g.,][]{kremer2019cmcgrid}.
We therefore limit our fitting procedure for each cluster to only the models with $R_g$ and $Z$ closest (in linear and logarithmic scales, respectively) to their observed present-day values, as reported by \citet{baumgardt2018mean} and \citet[][2010 edition]{harris1996catalog}, respectively.
Hence, for any individual cluster, we only optimize over $N$ and $r_v$. For simplicity, a constant tidal radius is assumed, although we caution that GC orbits in the Galaxy generally induce time-dependent tidal forces (including possible close pericenter passages). Furthermore, the modern-day distance of a GC to the Galactic center may not be representative of the average tidal force \citep{baumgardt2018mean}.

\subsection{Synthetic Observables and Cluster Fitting} \label{extractingclusterobservables}

In order to match an observed GC with a best-fit model on our grid, we identify models whose dynamical properties most closely match the observed cluster features.
The most direct dynamical observables of a GC are the surface brightness and velocity dispersion profiles, so we extract a simulated SBP and VDP from each model snapshot for comparison to the corresponding observed profiles.

%\kyle{This paragraph needs fixed a little bit. CMC actually records 1D positions and 2D velocities (I misspoke before). We create 3D values by drawing random angles. Then we translate these 3D values to 2D projections by again assuming a random projection. I suggest just saying something very simple}\textcolor{red}{Nicholas: Okay, I'll take care of this, just wanted to note that I don't use random numbers to get the 3D positions}
%Due to the spherical symmetry assumption of \texttt{CMC}, snapshot catalogs only contain radial positions and two-dimensional kinematic information of each star.
%All relevant spatial observables concern the cluster's projection on the plane of the sky.
%We convert the positions of stars recorded by \texttt{CMC} into these projected observables using the probability density function of projected two-dimensional radial distances given some three-dimensional radius.
%In particular, a star located a radial distance $r$ from the center of the cluster has a weight probability $p(a,b;r)$ of lying within projected radial distances $d=a$ and $d=b>a$, with $p$ given by 
Since GCs are observed only in projection on the sky, we project our simulated stellar positions and velocities onto a two-dimensional plane by assuming spherical symmetry.
In particular, a star with a three-dimensional radius $r$ has a probability $p(a,b;r)$ of lying within projected radial distances $d=a$ and $d=b>a$ given by

\begin{equation} \label{eqn:p}
    p(a, b; r) = \sqrt{1 - \min(a, r)^2/r^2} - \sqrt{1 - \min(b, r)^2/r^2}\mathrm{.}
\end{equation}

We calculate the surface brightness $\Sigma_V(a,b)$ and one-dimensional velocity dispersion $\sigma_v(a,b)$ for $80$ logarithmically spaced bins with an inner bin of $10^{-3}$ pc and an outer bin given by the maximum radial position of a star in the catalog.
For $\sigma_v(a,b)$, we only include evolved bright stars (\textsc{sse}/\textsc{bse} with startypes \texttt{2}-\texttt{9}) to mimic the use of bright stars in real VDP measurements \citep[e.g.,][]{kamann2017stellar,ferraro2018mikis}.

Given this two-dimensional distribution described by Equation \ref{eqn:p}, the average $V$-band surface brightness $\Sigma_V(a,b)$ in a projected radial bin bounded by $d\in(a,b]$ can be calculated as

\begin{equation}
    \begin{split}
        \Sigma_V&(a,b) = \\
        &-2.5\log_{10}\left( \frac{100\textnormal{ pc}^2\textnormal{ arcsec}^2}{\pi(b^2-a^2)}\sum_i\frac{p(a,b; r_i)}{10^{M_{V,i}/2.5}}\right) + A_V, \\
    \end{split}
\end{equation}

\noindent where $r_i$ and $M_{V,i}$ are the three-dimensional radial distance and absolute $V$-band magnitude of the $i$th star in the simulation, respectively, and $A_V$ is the $V$-band extinction.
For simplicity, all stars are assumed to be blackbodies, which should reasonably approximate their actual magnitudes, particularly for more massive stars where molecular lines are less prominent.
Stellar magnitudes thus take the form

\begin{equation} \label{eqn:bb}
M_{V,i} = -2.5\log_{10}\left(\frac{R_i^2}{(10\,\textrm{pc})^2F_{\textrm{ZP},\lambda}}\frac{\int f(\lambda,T_i)\mathcal{T}(\lambda)\mathrm{d}\lambda}{\int \mathcal{T}(\lambda)\mathrm{d}\lambda}\right),
\end{equation}

\noindent where $F_{\mathrm{ZP},\lambda}=3.57453\times10^{-9}$ erg cm$^{-2}$ s$^{-1}$ \AA$^{-1}${ }is the zero-point spectral flux, $f(\lambda,T)$ is the wavelength-space Planck distribution for temperature $T$, and $\mathcal{T}(\lambda)$ is the transmission function for the filter \citep{casagrande2014synthetic}.
Photometric magnitudes are derived using the generic Johnson $V$-band filter function and Vega zero point from the SVO Filter Profile Service, a public repository for astronomical filter parameters \citep{rodrigo2013filter,rodrigo2017svo}.
The $V$-band extinction $A_V$ of a GC is computed using the standard \citet{cardelli1989relationship} extinction law as $A_V=3.1\times E(B-V)$ where $E(B-V)$ is taken from the \citet[][2010 edition]{harris1996catalog} catalog.
While the blackbody approximation deviates significantly for cool M dwarfs which have significant molecular absorption lines \citep{allard1994influence,baraffe1995new}, the approximation is a reasonable estimate for brighter and hotter stars whose continuum emission dominates the profile.
Equation \ref{eqn:bb} is also applied in Section \ref{binary} to select binaries, though the cut is restricted to relatively bright main sequence stars where the blackbody approximation should be expected to hold.

The velocity dispersion $\sigma_v(a,b)$ in the same radial bin is given by

\begin{equation}
    \sigma_v(a,b) = \sqrt{\frac{\sum_i|\Vec{v}_i|^2p(a,b;r_i)}{3\cdot\sum_ip(a,b;r_i)}}, \\
\end{equation}

\noindent where $\vec{v}_i$ is the three-dimensional velocity vector of the $i$th star.
This expression for $\sigma_v(a,b)$ assumes an isotropic velocity dispersion.
Though in principle the tangential and radial components of the velocity dispersion may differ, in most cases the ratio of the two is very close to $1$\response{, especially near to the cluster center} \citep{watkins2015hubble}.
Furthermore, though some have claimed detection of coherent rotation within many GCs \citep[e.g.,][]{kamann2017stellar}, this subtle behavior is not captured in our models, and we do not consider \response{departures from spherical symmetry} in this work.
\response{Such rotation is, in any case, usually much smaller than the velocity dispersion, and should not be expected to change it significantly, especially in the central regions of the GC where \texttt{CMC} is expected to be most accurate.
While spherical asymmetry is the subject of very interesting work \citep[see, e.g.,][where measurements of Palomar 5's tidal tails are used in fitting]{gieles2021supra}, they lie beyond the scope of this work.}

We assess the relative likelihood that a given \texttt{CMC} model fits observed SBPs/VDPs by computing $\chi^2$ statistics between this data and linearly interpolated model SBPs/VDPs \citep{heggie2008monte,giersz2009monte,giersz2011monte,heggie2014mocca,kremer2018black,kremer2019initial}.
The fitness of a model with a given GC is assessed using $\tilde{\chi}_{\mathrm{SBP}}^2=\chi_{\mathrm{SBP}}^2/N_{\mathrm{SBP}}$ and $\tilde{\chi}_{\mathrm{VDP}}^2=\chi_{\mathrm{VDP}}^2/N_{\mathrm{VDP}}$, the $\chi^2$ statistics between the model SBP/VDP and the observations normalized by the number of observational data points.
For a given GC, we consider well-fitting snapshots to have \response{the fitting heuristic $s\equiv\max\left(\tilde{\chi}_{\mathrm{SBP}}^2,\tilde{\chi}_{\mathrm{VDP}}^2\right)<10$.}
Hence, to be a ``good fit'' to the data, a snapshot must be a reasonably good fit to both the SBP and VDP.
For diagnostic reasons, we also report for the best-fitting snapshot of each cluster $\tilde{\beta}_{\mathrm{SBP}}^2$ and $\tilde{\beta}_{\mathrm{VDP}}^2$, defined as the reduced $\chi^2$ sums with terms weighted by the sign of the residual.
These statistics parameterize the extent to which a given snapshot overestimates or underestimates a cluster's surface brightness or velocity dispersion, and are included to guide the creation of future models in order to better fit particular observed GCs.

The SBPs are taken from ground-based observations by \citet{trager1995catalogue}\footnote{As the publicly available SBP for NGC 2419 appears to be multi-valued, we restrict this profile to datasets in \citet{trager1995catalogue} which follow the branch shown in the plot in their paper, and estimate the SBP uncertainties using these points alone.}.
While other data sets such as the \citet{noyola2006surface} SBPs for $38$ GCs may better probe the surface brightnesses of GC cores, particularly for core-collapsed clusters, we opt to exclude this data in order to avoid assigning ad hoc relative weights between the Trager and Noyola profiles.
The VDPs are taken from the radial velocity measurements of \citet{kamann2017stellar}, \citet{ferraro2018mikis}, and \citet{baumgardt2018catalogue}, as well as the proper motion measurements of \citet{watkins2015hubble} and \citet{baumgardt2018mean}.
The VDPs measured using radial velocities and proper motions are generally observed to be consistent with one another for a given cluster\response{--VDP uncertainties across all data sets are thus taken as reported without any rescaling or homogenization.}
Thus, we fit to the combination of these VDPs for maximal constraint.

As \citet{trager1995catalogue} do not calculate formal uncertainties on their SBP measurements but instead provide a Chebyshev polynomial fit and coarse quality weights $w_i$ for each data point, we follow the procedure in \citet{mclaughlin2005resolved} to estimate uncertainties.
In particular, we assume that the measurement uncertainties are inversely proportional to $w_i$ and that the third-order Chebyshev polynomial fits have $\tilde{\chi}^2=1$ exactly.
We then estimate the uncertainty of the $i$th data point to be $\delta\Sigma_V=\delta\Sigma_{V,0}/w_i$, where $\delta\Sigma_{V,0}$ is estimated separately for each cluster as $\delta\Sigma_{V,0}=\sqrt{(N_{\mathrm{SBP}}-4)^{-1}\sum^N_{i=1}w_i^2e_{\Sigma_V,i}^2}$ where $e_{\Sigma_V,i}$ is the surface brightness residual of the $i$th data point with respect to the Chebyshev polynomial fit.
Since this effectively discards observations for which \citet{trager1995catalogue} assign $w_i=0$, we omit these points when normalizing $\tilde{\chi}_{\mathrm{SBP}}^2$ by $N_{\mathrm{SBP}}$.
As the publicly available SBP for NGC 2419 appears to be multi-valued, we restrict this profile to datasets in \citet{trager1995catalogue} which follow the branch shown in the plot in their paper, and estimate the SBP uncertainties using these points alone.

\begin{figure*}
    \centering
    \includegraphics[width=\textwidth]{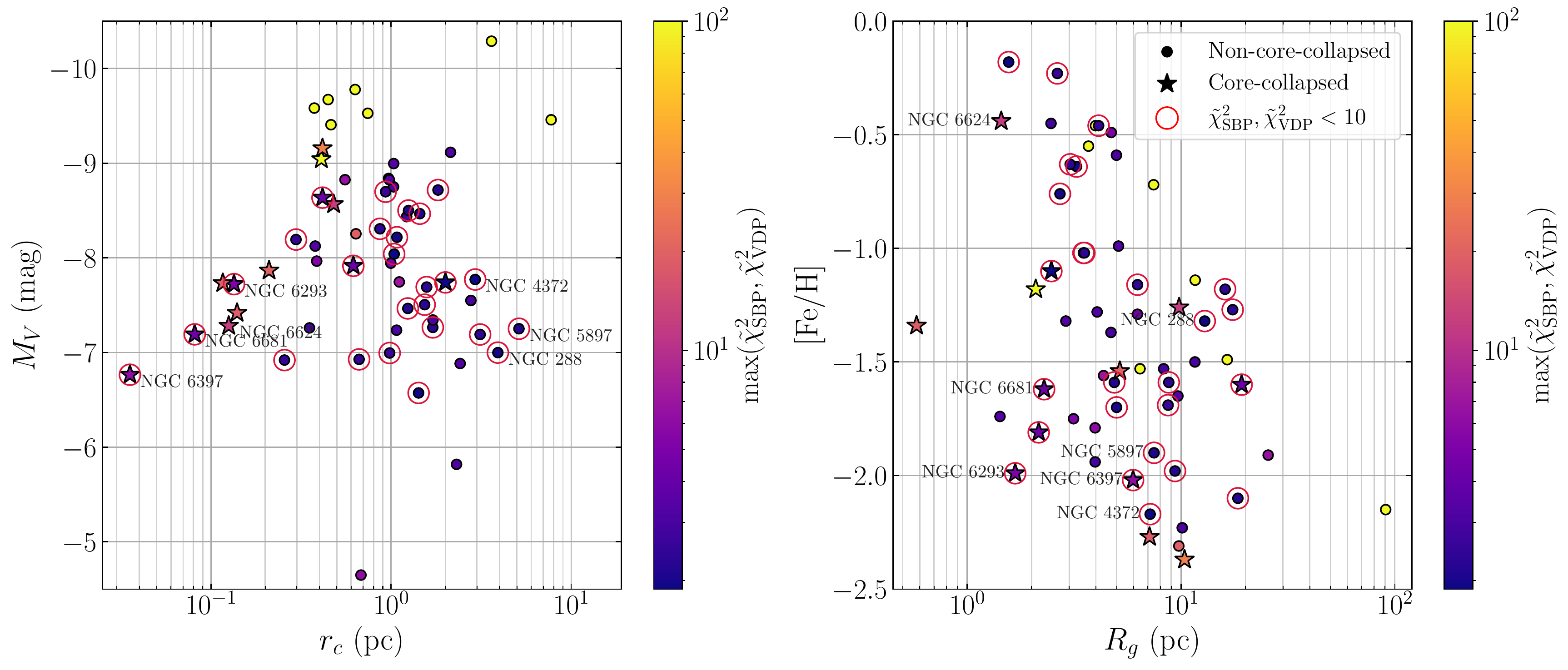}
    \caption{The GCs for which best-fit \texttt{CMC Cluster Catalog} models are identified, plotted in $r_c$--$M_V$ (\textit{left}) and $R_g$--$[\mathrm{Fe}/\mathrm{H}]$ (\textit{right}) space.
    GCs are color-coded by $s=\max\left(\tilde{\chi}_{\mathrm{SBP}}^2,\tilde{\chi}_{\mathrm{VDP}}^2\right)$, with some GCs saturating the color bar from above.
    Clusters which are considered ``well-fit'' ($s<10$) are circled in red, and clusters which are specifically discussed in the text are labeled.}
    \label{fig:summary_plot}
\end{figure*}

Figure \ref{fig:summary_plot} shows the GCs for which both the SBP and VDP are sampled with at least five points, in both the core radius--brightness ($r_c$--$M_V$) and galactocentric distance--metallicity ($R_g$--$[\mathrm{Fe}/\mathrm{H}]$) planes, with $M_V$, $r_c$, and $[\mathrm{Fe}/\mathrm{H}]$ taken from \citet[][2010 edition]{harris1996catalog} and $R_g$ from \citet{baumgardt2018mean}.
Points are color-coded by the fitting statistic $s$, with well-fit clusters circled.
Despite the wide range of Milky Way GC properties, we are able to satisfactorily fit a wide range of clusters across the observed parameter space.
As expected, well-fit clusters are concentrated at lower brightnesses, specifically at dimmer $M_V\gtrsim-9.5$, indicating a lack in grid coverage at larger masses.
Unsurprisingly, the quality of the fit does not obviously correlate with the present-day galactocentric distance or metallicity, as the bulk cluster dynamics are less sensitive to these parameters.

\subsection{Identification of Stellar Exotica} \label{exotica}

In the subsequent sections, we examine best-fitting models for seven specific GCs, during which we make comparisons to the observed population of low-mass X-ray binaries, millisecond pulsars, and cataclysmic variables.
Although we do not use these stellar exotica as factors in our goodness of fit measurements due to their large uncertainties, we can use the rough numbers as guideposts to further constrain and explore our models.

An X-ray binary (XRB) is a mass-transferring binary where the donor, typically a main-sequence star, accretes onto a compact object, either a neutron star or a black hole.
While short-lived high-mass X-ray binaries (with \response{OB-type} donors) dominate the X-ray sources in young, star-forming populations, low-mass X-ray binaries (LMXBs) are believed to form in the dense cores of GCs \citep{pooley2003dynamical,fabbiano2006populations}.
In our models, we consider as XRBs any main-sequence star (\textsc{sse} startype $0$--$1$) in a mass-transferring binary with a neutron star or black hole.
Our models contain characteristically between $0$ and a few XRBs.

One possible outcome of a disrupted LMXB is a millisecond pulsar.
Millisecond pulsars (MSPs) are rapidly rotating pulsars with periods on the order of milliseconds.
Unlike standard pulsars, which are young neutron stars rotating fast enough to beam, MSPs are older neutron stars which have been ``recycled'' by accretion from a companion.
Whereas the former have periods $P\simeq0.1$--$3$ s, mass transfer onto the latter spins MSPs up to periods as small as $1.5$ ms \citep{lorimer2008binary}.
Such objects are expected to be formed at an enhanced rate in the high-density environments at the center of GCs, particularly in core-collapsed clusters with very few black holes \citep{claire2019millisecond}.
In our models, we identify MSPs as neutron stars with periods $P<10$ ms.

Cataclysmic variables (CVs) are usually low-mass main-sequence stars undergoing mass transfer onto a white dwarf via Roche lobe overflow.
As their name suggests, they are characterized by large, often rapid flux variability, in some cases undergoing violent eruptions in the form of novae or dwarf novae \citep{robinson1976structure}.
GCs are believed to harbor significant CV populations which can help guide study into their evolution as well as possible role as SNe Ia progenitors \citep{ivanova2006formation,knigge2012cvs,maoz2014observational}.
Similarly to XRBs, we identify CVs as \textsc{sse} startype \texttt{0} stars undergoing mass transfer onto white dwarfs.
Our models contain a wide range of CVs spanning from between a few and $\sim100$ CVs.
Interestingly, because most CVs originate from primordial binary progenitors, the number of CVs in a GC actually correlates inversely with its central density.
This scaling is in contrast to that of other objects (e.g., MSPs) whose formation is predominantly dynamical.

\begin{figure*}
    \centering
    \includegraphics[width=\textwidth]{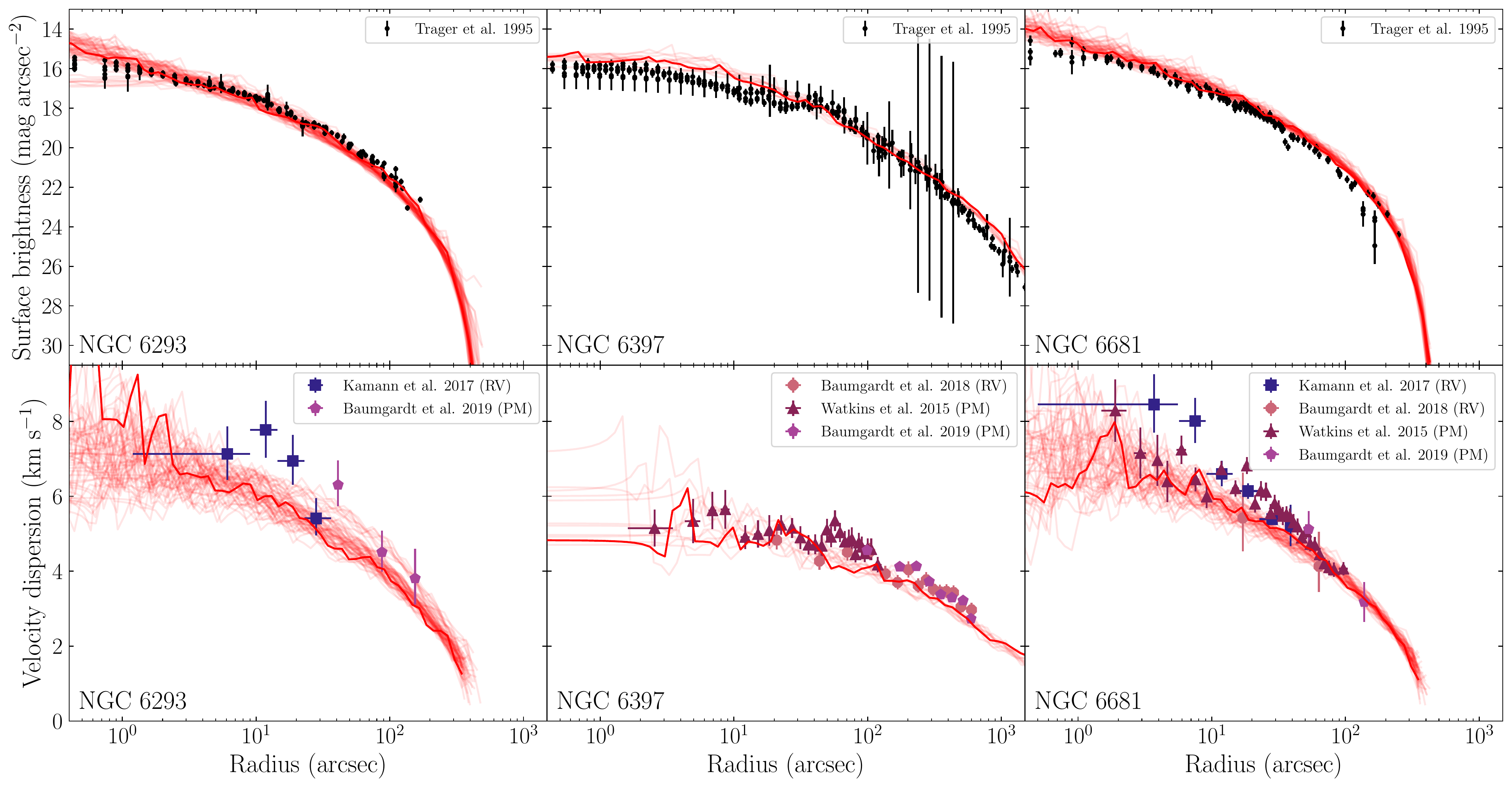}
    \caption{SBPs and VDPs together with best-fitting models for the core-collapsed clusters NGC 6293, NGC 6397, and NGC 6681. The best-fit profile is shown as an opaque red curve, and the SBPs and VDPs of well-fitting snapshots ($s<10$) are shown as translucent red curves.}
    \label{fig:profile_cc}
\end{figure*}

\begin{figure*}
    \centering
    \includegraphics[height=0.95\textheight]{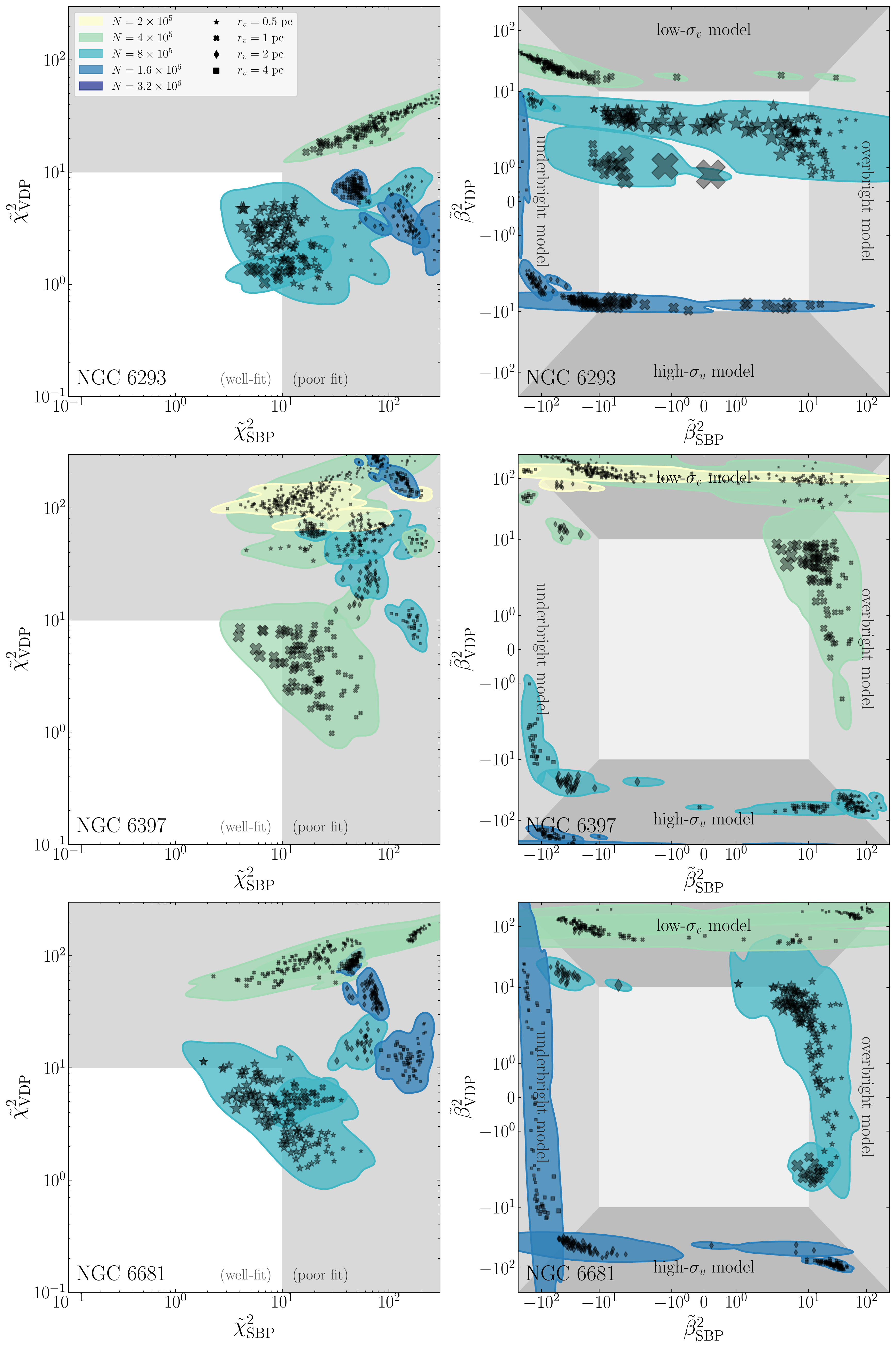}
    \caption{\textit{Left:} Plots of $\tilde{\chi}_{\mathrm{SBP}}^2$ and $\tilde{\chi}_{\mathrm{VDP}}^2$ for simulated SBPs and VDPs of snapshots with $t\geq10$ Gyr against NGC 6293, NGC 6397, and NGC 6681, which are all core-collapsed today.
    The colorful, shaded regions are level curves of kernel density estimates for individual models to guide the eye, and are colored by the initial number of stars in the simulation.
    We consider ``good fits'' to be given by $s<10$.
    \textit{Right:} Same as the left plots, except for $\tilde{\beta}_{\mathrm{SBP}}^2$ and $\tilde{\beta}_{\mathrm{VDP}}^2$.
    These parameters are estimators for how much a model overestimates the SBP or VDP of a cluster.
    Points far from the origin are reduced in size for clarity.}
    \label{fig:chisqplot_cc}
\end{figure*}

Blue straggler stars are unusually bright/blue main sequence stars which have been rejuvenated via either accretion from another star \citep[e.g.,][]{chen2004effects} or a collision sometime in the GC's history \citep[e.g.,][]{glebbeek2008evolution}.
Coupling to both standard binary evolution and cluster dynamics, the radial distribution of blue stragglers within GCs has gained significant attention as a tracer of a cluster's dynamical history \citep[e.g.,][]{ferraro2012dynamical,ferraro2018hubble}.
We leave the modeling of these interesting sources to a future work.

\section{Fitting Clusters to Observations} \label{fitting}

\begin{figure*}
    \centering
    \includegraphics[width=0.95\textwidth]{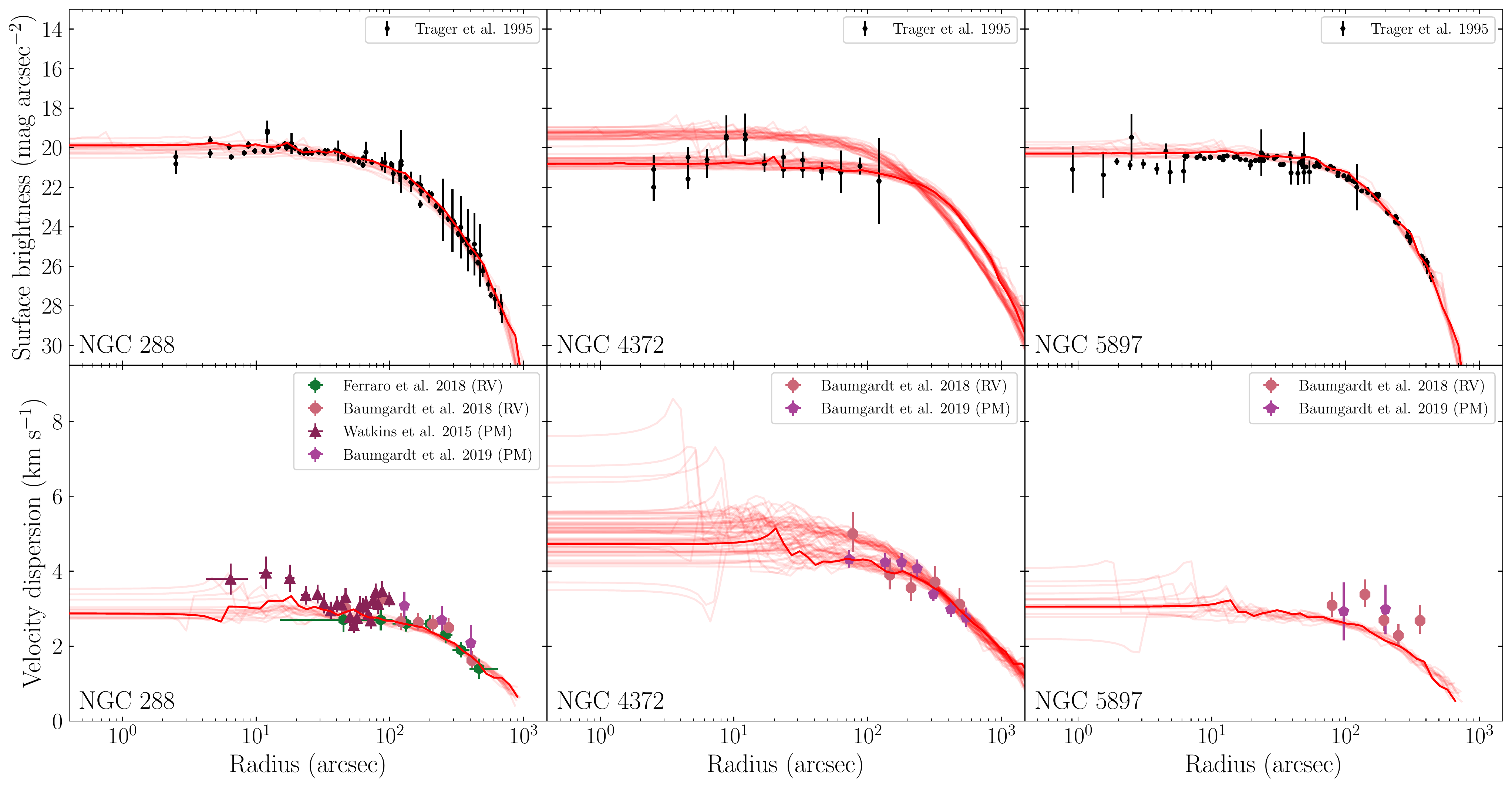}
    \caption{Same as Figure \ref{fig:profile_cc} but for non-core-collapsed clusters NGC 288, NGC 4372, and NGC 5897.}
    \label{fig:profile_ncc}
\end{figure*}

\begin{figure*}
    \centering
    \includegraphics[height=0.95\textheight]{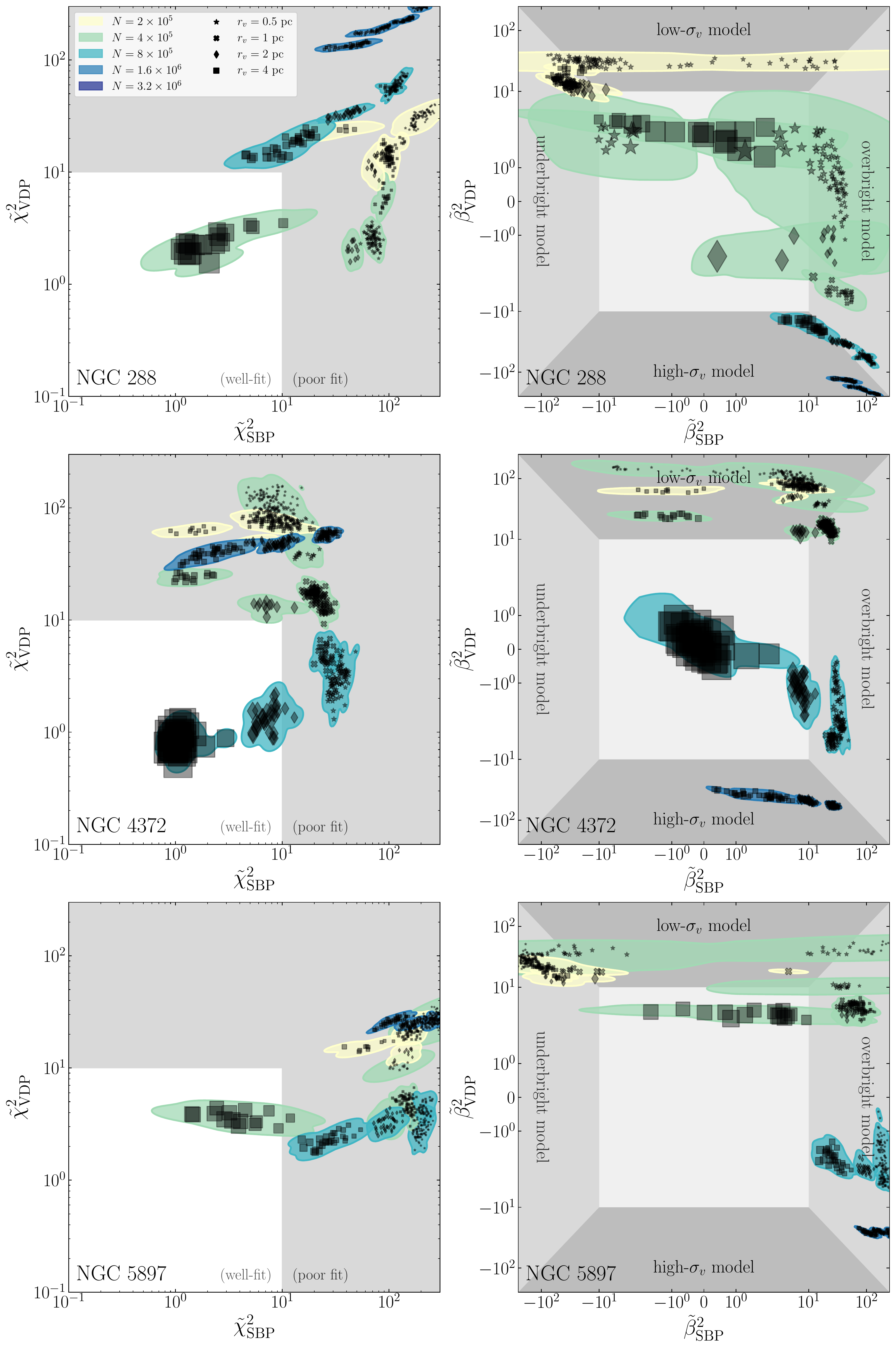}
    \caption{Same as Figure \ref{fig:chisqplot_cc} but for the non-core-collapsed clusters NGC 288, NGC 4372, and NGC 5897, for all of which well-fitting snapshots exist.}
    \label{fig:chisqplot_noncc}
\end{figure*}

Using the procedure described in Section \ref{extractingclusterobservables}, we identify best fits for $59$ Milky Way GCs which have $N_{\mathrm{SBP}},N_{\mathrm{VDP}}\geq5$ (see Appendix \ref{appendix}).
We obtain at least one ``well-fitting'' snapshot ($s<10$) for $26$ GCs, including the core-collapsed clusters NGC 6293, 6397, and 6681 (Figures \ref{fig:profile_cc} and \ref{fig:chisqplot_cc}), and non-core-collapsed clusters NGC 288, 4372, and 5897 (Figures \ref{fig:profile_ncc} and \ref{fig:chisqplot_noncc}).
\response{We emphasize that these best-fitting snapshots exist on the \texttt{CMC Cluster Catalog} as-is, without interpolation or any directed effort to fit any particular cluster.
A more precise representation of a particular GC requires the creation of new \texttt{CMC} models.}
These GCs cover a range of dynamical states, masses, distances, and metallicities, and allow us to benchmark our model predictions with a variety of different cluster properties.
\response{The non-core-collapsed clusters chosen for further examination have large core radii (as seen from Earth)---together with the core-collapsed clusters of interest, these GCs allow us to explore the full range of realistic values of $r_c$.
However, we note that some GCs with intermediate core radii (e.g., NGC 6352) are also fit well by some models in the \texttt{CMC Cluster Catalog}, as is apparent in Figure \ref{fig:summary_plot}.}

\response{Additionally, t}o demonstrate the ability to straightforwardly supplement the \texttt{CMC Cluster Catalog} to fit new clusters \response{and to present one example of a model refinement for a specific GC}, we also extend the model grid in Section \ref{ngc6624} to fit NGC 6624, an interesting high-metallicity cluster known for its high-energy emission and large number of recorded millisecond pulsars.

\subsection{NGC 6293} \label{ngc6293}

Reaching about $0.5$ kpc at perigalacticon, NGC 6293 is a low-metallicity, core-collapsed GC which lies very close to the Galactic center---an intense tidal environment which spatially flattens the cluster's shape \citep{chen2010morphological,baumgardt2018mean}.
Within the \texttt{CMC Cluster Catalog}, we locate $59$ well-fitting snapshots from two core-collapsed models, \textsc{n8-rv1-rg2-z0.01} ($8$ snapshots) and \textsc{n8-rv0.5-rg2-z0.01} ($51$ snapshots), both with initial $N=8\times10^5$.
While the SBPs are generally matched quite well within the cores, some snapshots slightly overestimate the core brightness and some underestimate it. Hence well-fitting snapshots should produce predictions for cluster properties (e.g., total mass) which ``surround'' their true values.
While all well-fitting snapshots underestimate the VDP somewhat, observational uncertainties are comfortably large enough to be consistent with predictions.

NGC 6293 is associated with at least one soft X-ray source \citep[XTE J1709-267,][]{jonker2003chandra}.
Of the $59$ well-fitting snapshots, $54$ snapshots do not have any XRBs and $5$ snapshots contain a single XRB ($3$ from the $r_v=1$ pc model pc and $2$ from the $r_v=0.5$ pc model).
Furthermore, all of these XRBs have low-mass donors with \textsc{sse} startype \texttt{0} ($M\lesssim0.7$ $M_\odot$).
% (defined in our models to be some low-mass main-sequence star, \textsc{sse} startype \texttt{0} accreting onto a neutron star or black hole).
Despite possible observational biases and incompleteness on the total number of XRBs in the cluster (especially for a cluster near the Galactic Center), our models appear consistent with the ability for NGC 6293 to produce a small number of X-ray sources.

\subsection{NGC 6397} \label{ngc6397}

NGC 6397 is a nearby \citep[$D=2.44\pm0.04$ kpc,][]{baumgardt2018mean} metal-poor, core-collapsed GC whose close proximity has attracted significant study of its white dwarf and low-mass stellar populations \citep{paresce1995very,cool1996main,taylor2001helium,hansen2007white}.
In addition, NGC 6397 has $15$ known CV candidates, a quiescent LMXB (qLMXB), and $1$--$2$ MSPs \citep{cool1995discovery,grindlay2001chandra,cohn2010identification,dieball2017far}.
We locate $11$ well-fitting snapshots from the model \textsc{n4-rv1-rg8-z0.01}.
While all of these snapshots tend to overestimate the surface brightness and underestimate the velocity dispersion somewhat, they do so within our tolerances.
By eye, it appears that the slight overestimation of the SBP occurs primarily at $\approx5$ pc, the model is slightly brighter than the data.

The well-fitting snapshots for this cluster each have between $11$ and $13$ CVs.
%binaries consisting of low-mass main sequence stars (\textsc{sse} startype \texttt{0}) accreting onto white dwarfs, which roughly matches the observed number of CVs only if most are active enough to be detected.
In similar agreement with observations, these snapshots also have between $1$ and $2$ XRBs each (all of which have $M\lesssim0.7$ $M_\odot$ donors).
Finally, these snapshots all contain $2$ MSPs.
%, defined to be neutron stars in our models with periods $T<10$ ms.
Of course, while these small numbers should not be taken as precise predictions, they provide a measure of reassurance that our models generate these populations in reasonable numbers.

\subsection{NGC 6681} \label{ngc6681}

NGC 6681 (M70) is a core-collapsed cluster which has occasionally been used to derive distortion solutions for instruments operating in the far-ultraviolet \citep[e.g.,][]{sohn2018new}.
Like NGC 6293, NGC 6681 lies quite close to the Galactic center \citep[$R\sim0.8$ kpc,][]{baumgardt2018mean} and observations also suggest some degree of tidal deformation \citep{han2017spatial}.
We find $49$ well-fitting snapshots from the core-collapsed model \textsc{n8-rv0.5-rg2-z0.01}.
As this particular model provides snapshots consistent with the SBP and VDP of NGC 6293 as well, one can view NGC 6681 as somewhat similar to NGC 6293 from a dynamical perspective, too.
As with NGC 6293, all snapshots slightly underestimate the VDP, but within tolerance of observational uncertainties.

\subsection{NGC 288} \label{ngc288}

A relatively metal-rich, non-core-collapsed GC, NGC 288 received much attention in the 1990s and early 2000s as its similar metallicities and distances to those of NGC 362 provided promising avenues to constraining the age difference between the two clusters \citep{green1990population,sarajedini1990new,bellazzini2001age}.
There have since been a large number of studies into its dynamics \citep[e.g.,][]{piatti2018extended} and stellar populations \citep[e.g.,][]{roh2011two}.
Within our model grid, we identify $16$ well-fitting snapshots from a large $r_v=4$ model, \textsc{n4-rv4-rg8-z0.1}.
Both the SBPs and VDPs from these snapshots fit observations exceptionally well.

Using Chandra, \citet{kong2006chandra} report between $2$ and $5$ possible CVs or other chromospherically active binaries within the half-mass radius of NGC 288.
Meanwhile, our models predict between $27$ and $33$ CVs.
This discrepancy may result from the high temporal variability in the activity of CVs, which could make quiescent accreting binaries difficult to detect, or some other source of observational incompleteness.

\subsection{NGC 4372} \label{ngc4372}

While \citet{trager1995catalogue} do measure the $V$-band SBP for this cluster, they do not report data outside the $1.75$ arcmin core radius reported by \citet[2010 edition]{harris1996catalog}.
Hence, the SBP neither contains the characteristic turnover in brightness nor constrains particularly well the size of the cluster's core.
Therefore, while the \texttt{CMC Cluster Catalog} includes $30$ well-fitting snapshots each from $N=8\times10^5$ models \textsc{n8-rv4-rg8-z0.01} and \textsc{n8-rv2-rg8-z0.01}, the initial virial radius of the cluster is largely uncertain.
Though the snapshots come from models with different initial $r_v$, their similar initial masses and dynamical states produce a relatively small spread in the predicted mass of the cluster.
\citet{kacharov2014study} estimate a cluster mass $M=(2.0\pm0.5)\times10^5$ $M_\odot$, compatible with the estimated mass from snapshots at the $1\sigma$ level.
They also estimate for this cluster a mass-to-light ratio between $1.4$ and $2.3$, which is broadly compatible with predicted values for all seven of our clusters of interest (see Section \ref{mml}).

Within NGC 4372, \citet{kaluzny1993contact} identify a candidate CV, though atmospheric effects imply a large amount of incompleteness in the data.
Using X-ray observations from XMM-Newton, \citet{servillat2008xmm} were unable to detect specific CVs---although they are limited by X-ray luminosity---although they report unresolved emission consistent with a population of $\sim20$ CVs.
As our well-fitting snapshots come from two models with different initial $r_v$, we find a bimodal distribution with modes approximately around $26$ and $84$ CVs, with well-fitting snapshots having between $21$ and $98$.
The lower and higher modes correspond to snapshots from the $r_v=2.0$ pc and $r_v=4.0$ pc models, respectively.
Intuitively, this correlation between the number of CVs and $r_v$ (inversely related to the central density) is due to the origination of most CVs in primordial binaries, which are less likely to disrupt in a less dense cluster \citep{kremer2019cmcgrid}.
The observed number of CVs is consistent with the predicted number from the $r_v=2$ pc model.

\subsection{NGC 5897} \label{ngc5897}

NGC 5897 is a non-core-collapsed cluster at a distance $D\approx12.6$ kpc \citep{baumgardt2018mean}.
While the VDP has been measured using both radial velocities \citep{baumgardt2018catalogue} and proper motions \citep{baumgardt2018mean}, these measurements have not extended into the core of the cluster, and thus do not particularly well constrain the dark mass distribution there.
We find $14$ well-fitting snapshots in the model \textsc{n4-rv4-rg8-z0.01}, which differs from the well-fitting model for NGC 288 only in that its metallicity is lower.
Hence, between the two clusters, one expects similarity in their dynamics but not necessarily their stellar populations.

\subsection{NGC 6624} \label{ngc6624}

One of a handful of clusters with $\gamma$-ray emission $>100$ MeV \citep{tam2011gamma}, NGC 6624 is an interesting, high-metallicity \citep[$\lbrack\mathrm{M}/\mathrm{H}\rbrack\approx-0.44$;][2010 edition]{harris1996catalog} GC which is known to contain at least $4$ MSPs in addition to $2$ young pulsars \citep{biggs1994two,lynch2012timing}.
In the past, it has been argued that the relatively large spin period derivatives of an MSP near the cluster center is evidence for an IMBH \citep{peuten2014puzzling,perera2017evidence}, though these signals have since been found to be consistent with dynamical interactions alone \citep{gieles2018mass,baumgardt2019no}.
Additionally, the cluster is known to contain at least one well-studied LMXB \citep[4U 1820-30;][]{biggs1994two}.
Importantly, the cluster lacks any well-fitting snapshots from the unmodified \texttt{CMC Cluster Catalog}---the best-fitting snapshot belongs to the model \textsc{n8-rv0.5-rg2-z1.0} for which $\tilde{\chi}_{\mathrm{SBP}}^2=11.95$.
We present NGC 6624 as a test case for introducing new models in order to fit a known SBP and VDP which is not satisfactorily fit by the main \texttt{CMC Cluster Catalog}.

\begin{deluxetable*}{l l l | c c c c | l l l l}
\centerwidetable
\tabletypesize{\normalsize}
\tablecolumns{15}
\tablewidth{0pt}
\tablecaption{New Models for NGC 6624\label{tab:newmodels}}
\tablehead{ \colhead{Model} & \colhead{$N_{\mathrm{snap}}$} & \colhead{$N_{\mathrm{good}}$} & \colhead{$r_v$} & \colhead{$R_g$} & \colhead{$[M/H]$} & \colhead{$N$} & \colhead{$\tilde{\chi}_{\mathrm{SBP}}^2$} & \colhead{$\tilde{\chi}_{\mathrm{VDP}}^2$} & \colhead{$\tilde{\beta}_{\mathrm{SBP}}^2$} & \colhead{$\tilde{\beta}_{\mathrm{VDP}}^2$}\\
\colhead{} & \colhead{} & \colhead{} & \colhead{pc} & \colhead{kpc} & \colhead{} & \colhead{$\times10^5$} & \colhead{} & \colhead{} & \colhead{} & \colhead{} }
\startdata
\textsc{n7-rv0.5-rg2-z1.0} & $570$ & $0$ & $0.5$ & $2$ & $0$ & $7$ & $14.51$ & $2.00$ & $6.23$ & $0.78$ \\
%\textsc{n9-rv0.5-rg2-z1.0} & $570$ & $0.5$ & $2$ & $1.0$ & $9$ & \textcolor{red}{what} & \textcolor{red}{what} & \textcolor{red}{what} & \textcolor{red}{what} \\
\textsc{n8-rv0.7-rg2-z1.0} & $437$ & $0$ & $0.7$ & $2$ & $0$ & $8$ & $12.24$ & $3.37$ & $9.11$ & $-2.70$ \\
\textsc{n7-rv1-rg2-z0.35} & $306$ & $13$ & $1$ & $2$ & $-0.46$ & $7$ & $5.22$ & $2.41$ & $3.04$ & $-2.17$ \\
\textsc{n6-rv1-rg2-z1.0} & $233$ & $0$ & $1$ & $2$ & $0$ & $6$ & $11.38$ & $1.76$ & $7.83$ & $-0.15$ \\
\textsc{n7-rv1-rg2-z1.0} & $218$ & $0$ & $1$ & $2$ & $0$ & $7$ & $13.60$ & $4.05$ & $10.94$ & $-3.90$ \\
\textsc{n9-rv1-rg2-z1.0} & $317$ & $0$ & $1$ & $2$ & $0$ & $9$ & $21.66$ & $13.20$ & $21.35$ & $-13.20$ \\
\hline
\enddata
\tablecomments{Parameters of additional models generated to better fit NGC 6624, alongside goodness-of-fit measures for closest-fitting snapshots.}
\end{deluxetable*}

Without the addition of any other models, the \texttt{CMC Cluster Catalog} already provides a satisfactory fit for $26$ out of $59$ of the GCs for which we attempt to locate an analogous model.
Of the remaining GCs, there are broadly three reasons why a GC may not be fit well by the \texttt{CMC Cluster Catalog}.
First, the GC's parameters may lie entirely outside of the range of parameter space probed by the \texttt{CMC Cluster Catalog}, although further simulations outside of this parameter space (e.g., new models with larger $N$ than the largest $N$ on the grid) may fit such clusters.
Second, the GC may lie between points in parameter space sampled by the grid; for example, a GC's initial $r_v$ may lie between $1$ and $2$ pc, but not especially close to either.
In this case, the inclusion of models which increase the resolution of the grid would fit these clusters.
Finally, the GC may differ from the models in other parameters besides those which have been varied in the grid, namely, $r_v$, $R_g$, $Z$, and $N$.
Such parameters include the initial binary fraction and initial mass function, which are not varied on the \texttt{CMC Cluster Catalog} but can generally be varied in \texttt{CMC} to potentially improve GC fits.

%A GC may lack a well-fitting snapshot on the main \texttt{CMC Cluster Catalog} for a number of reasons.
%First, the GC's parameters may lie out of the parameter space probed by the grid, for example if the GC is more massive than any model snapshot on the grid.
%Second, the GC may not be captured by the parameter grid, which is coarse over all four varied parameters--its initial $N$ may have laid, for example, in between $N=4\times10^5$ and $N=8\times10^5$ but especially not close to either.
%Finally, the GC may differ from the models in other parameters besides those which have been varied in the grid, namely, $r_v$, $R_g$, $Z$, and $N$.
NGC 6624 does not have a well-fitting snapshot according to $s<10$.
However, as its closest-fitting snapshots on the main grid do not appear to deviate from the observed data substantially, it likely lies within the second aforementioned category: the lack of good fits probably is due to the coarseness of the grid rather than its limited range.
Nevertheless, its closest-fitting snapshots provide guidance as to the manner in which the grid ought to be extended to obtain a good fit.
Guided by these pre-existing models, we supplement the model grid to better fit NGC 6624 in order to demonstrate targeted cluster fitting with further \texttt{CMC} simulations.

%Though NGC 6624 lacks well-fitting snapshots on the main model grid, its closest-fitting snapshots provide guidance as to the manner in which the grid ought to be extended to fit it well.
%As its closest-fitting snapshots on the main grid do not appear to deviate from the observed data substantially, it is likely that the lack of good fits is due to the coarseness of the grid rather than its limited range.

Notably, $\tilde{\chi}_{\mathrm{SBP}}^2=11.95$ for this cluster, indicating that some model SBPs are only slightly discrepant with the observed SBP.
Upon inspection, this disagreement arises because the model
The SBPs' core brightnesses are higher than observed while the outer halo brightnesses are slightly lower than observed.
Modifying initial $N$ can narrow these discrepancies (either directly by altering visible mass or indirectly by changing the relaxation time), as can modifying
$r_v$ \citep[which has been shown to influence progress toward core-collapse, see][]{kremer2018black}.

To better fit NGC 6624, we accordingly run $6$ new models and examine the extent to which they improve (or fail to improve) the fit (Table \ref{tab:newmodels}).
We consider separately a decrease in initial $N$ to $7\times10^5$ particles (\textsc{n7-rv0.5-rg2-z1.0}), and an increase in the virial radius to $r_v=0.7$ pc (\textsc{n8-rv0.7-rg2-z1.0}).
We also consider an increase in the virial radius to $r_v=1.0$ pc under three distinct initial particle counts, $N=(6,7,9)\times10^5$, (\textsc{n6-rv1-rg2-z1.0}, \textsc{n7-rv1-rg2-z1.0}, and \textsc{n9-rv1-rg2-z1.0}, respectively).
Finally, for $N=7\times10^5$, we also consider a decrease in metallicity to $Z=0.35\,Z_\odot$ (\textsc{n7-rv1-rg2-z0.35}), the observed metallicity reported by \citet[2010 edition]{harris1996catalog}.
As the metallicity is coarsely rounded up to $Z=1.0$ $Z_\odot$ in the main \texttt{CMC Cluster Catalog}, we consider this latter model to ascertain whether or not the structure of the GC displays fine sensitivity to $Z$.

\begin{figure}
    \centering
    \includegraphics[width=0.45\textwidth]{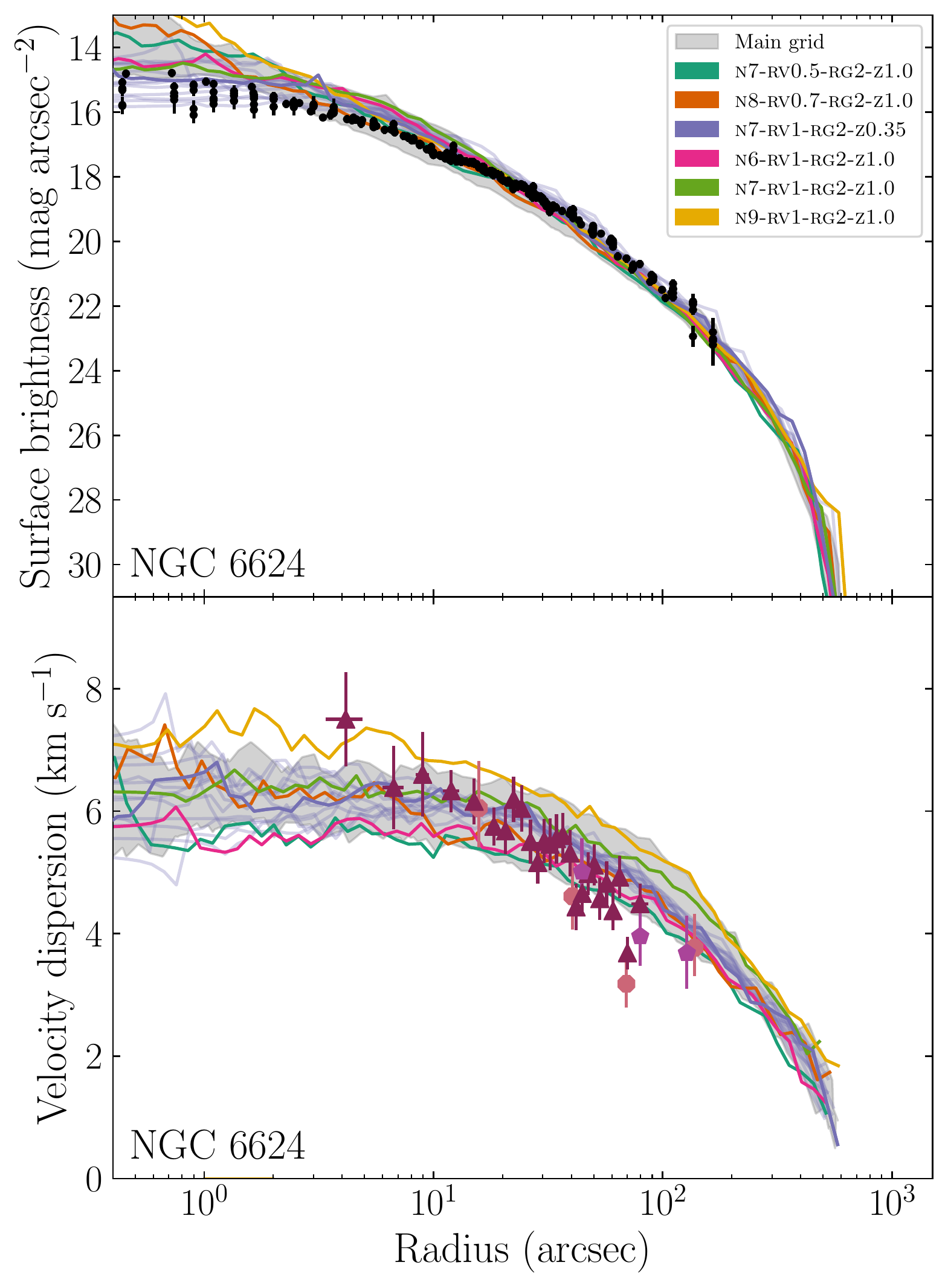}
    \caption{Same as Figures \ref{fig:profile_cc} and \ref{fig:profile_ncc} but for NGC 6624.
    As this cluster lacks any well-fitting models from the main \texttt{CMC Cluster Catalog}, we show the values spanned by the $30$ best-fitting snapshots from the main \texttt{CMC Cluster Catalog} in the gray region.
    The best-fitting snapshot from each of the new models is shown, as well as all well-fitting snapshots from the \textsc{n7-rv1-rg2-z0.35} model.}
    \label{fig:profile_bad_aug}
\end{figure}

\begin{figure*}
    \centering
    \includegraphics[width=\textwidth]{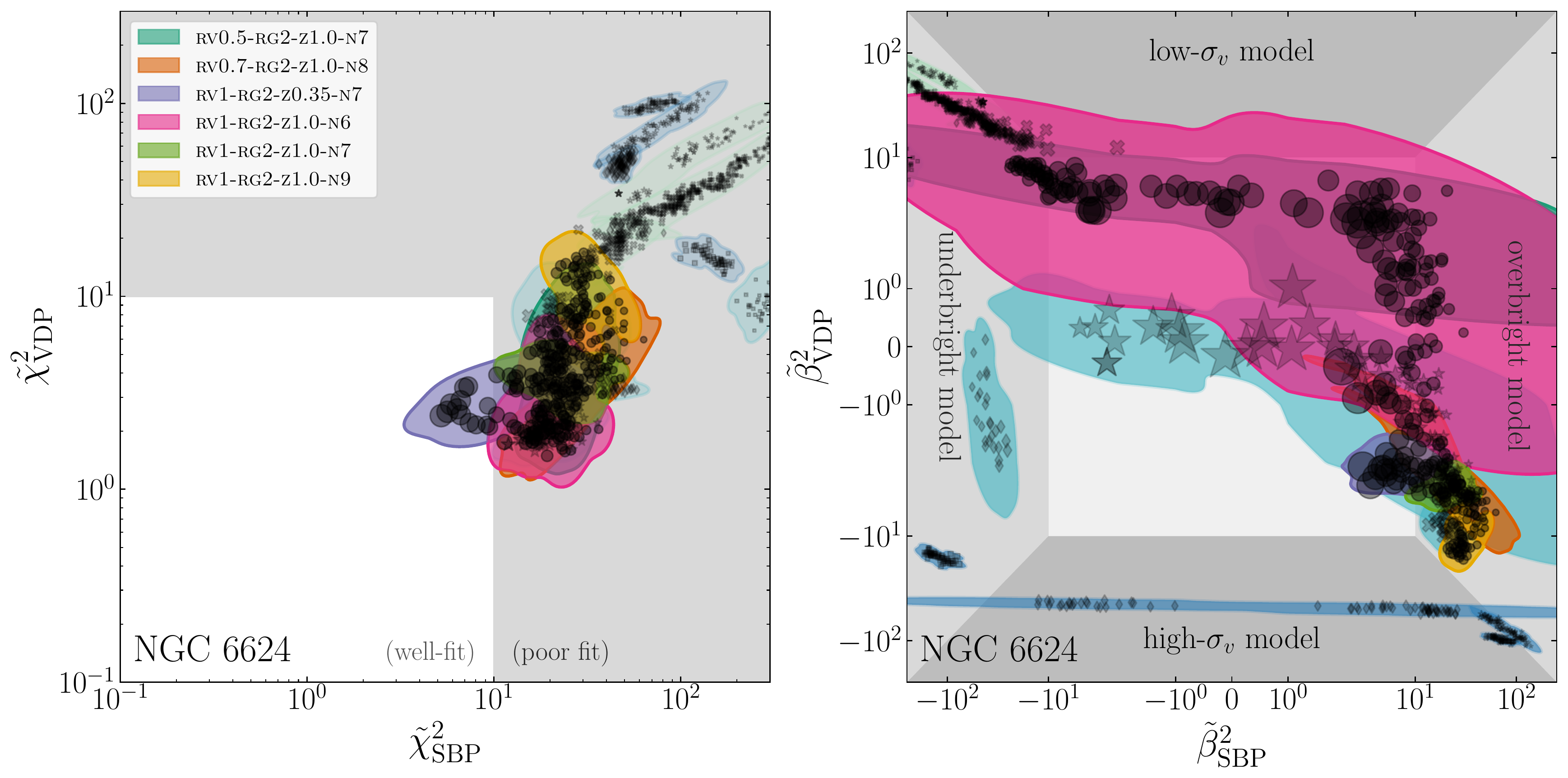}
    \caption{Same as Figure \ref{fig:chisqplot_cc} and \ref{fig:chisqplot_noncc}, but for NGC 6624 with the fitting metrics of additional models shown.
    Snapshots from models in the main \texttt{CMC Cluster Catalog} are shown in washed-out colors for reference.}
    \label{fig:chisqplot_poor_augment}
\end{figure*}

Of these $6$ models, we find that only \textsc{n7-rv1-rg2-z0.35} provides any well-fitting snapshots ($13$, Figures \ref{fig:profile_bad_aug} and \ref{fig:chisqplot_poor_augment}).
Within the other models, the best-fitting snapshots from \textsc{n7-rv0.5-rg2-z1.0}, \textsc{n8-rv0.7-rg2-z1.0}, and \textsc{n9-rv1.0-rg2-z1.0} all contain the same central overdensity as the best-fitting snapshots of the \texttt{CMC Cluster Catalog} proper.
%, suggesting that initial $r_v\approx1$ pc appears to be marginal for stalling core collapse to its present state by the present day. \newlin{The wording here is a bit awkward. More importantly, the conclusion is confusing. Only one of these three models has $r_v=1\,\rm{pc}$. $Z$ is the only parameter which is the same across these three models and different from the well-fitting one above. To me, the $Z$ difference is pretty revealing and important to discuss. Lower $Z$ increases the masses of central BHs, which we know from my first couple papers will in turn puff up the cluster. As a result, lower $Z$ should reduce the central brightness, which is exactly what you see.}
%\textcolor{red}{Indeed, our choice of $Z=0.35$ $Z_\odot$ (versus $Z=1.0$ $Z_\odot$) in the best-fitting model to better match the observed metallicity of the cluster appears to have decreased the model central density enough to achieve a good fit.
%Since $Z$ anticorrelates to the masses of the black hole }
The models \textsc{n7-rv1-rg2-z1.0} and \textsc{n6-rv1-rg2-z1.0} appear to decrease the core overdensity in the models, although they appear to be slightly overbright at $r_v\approx6$ arcsec $\approx0.2$ pc.% \newlin{Why is \textsc{n9-rv1.0-rg2-z1.0} (red) so much more centrally-dense than these two models (orange and yellow), which practically lie on top of one another but have only slightly smaller N. Is it just stochastic noise in the core between snapshots?}\claire{I think it's because N9 is much more massive than N7 and N6. And the N8 line is between N9 and N7/N6, which makes sense.}

Of the $13$ snapshots that fit NGC 6624 well, all lack MSPs.
Most of these snapshots also lack any XRBs, with $1$ snapshot containing a single XRB and $2$ snapshots containing $2$ XRBs.
However, $16$ additional snapshots pass the slightly relaxed fitting criterion $s<15$, with $6$ from different models: \textsc{n6-rv1-rg2-z1.0} (with $1$ MSP and $2$ to $4$ XRBs), \textsc{n7-rv1-rg2-z1.0} (with $2$ MSPs and $1$ XRBs), and \textsc{n8-rv0.7-rg2-z1.0} ($4$ to $5$ MSPs and $0$ to $1$ XRBs).
Our models are consistent with NGC 6624's single observed LMXB.
Although we find fewer MSPs in our well-fit models, the difference is small, and may be explained by moderate sensitivity of this number to the initial size, compactness, and metallicity of the cluster.
Since current observations aren't necessarily complete, this could also suggest that our models may somewhat under-produce these types of stellar exotica. However, it is also important to keep in mind that NGC 6624 exists in a more complex high-metallicity regime and lies behind a moderate amount of extinction. Coupled with uncertainty in the distance from Gaia \citep[$D=7.19\pm0.37$ kpc,][]{baumgardt2018mean} and initial mass function \citep[which may play a significant role;][]{weatherford2101black}, it remains plausible that both observational and modeling uncertainties could account for the deficit in MSPs and XRBs in the best-fitting models.

\section{Comparison to Cluster Properties}

In recent years, new observational surveys have led to the measurement of a breadth of physical observables across a broad sample of GCs, particularly those in the Milky Way.
For each of the seven GCs described in Section \ref{fitting}, we explore model predictions for cluster masses and mass-to-light ratios (Section \ref{mml}), binary fractions (Section \ref{binary}), mass segregation (Section \ref{massseg}), and black hole content (Section \ref{bh}), benchmarking these properties to observations whenever available.

\subsection{Cluster Masses and Mass-to-Light Ratios} \label{mml}

\begin{figure}
    \centering
    \includegraphics[width=0.45\textwidth]{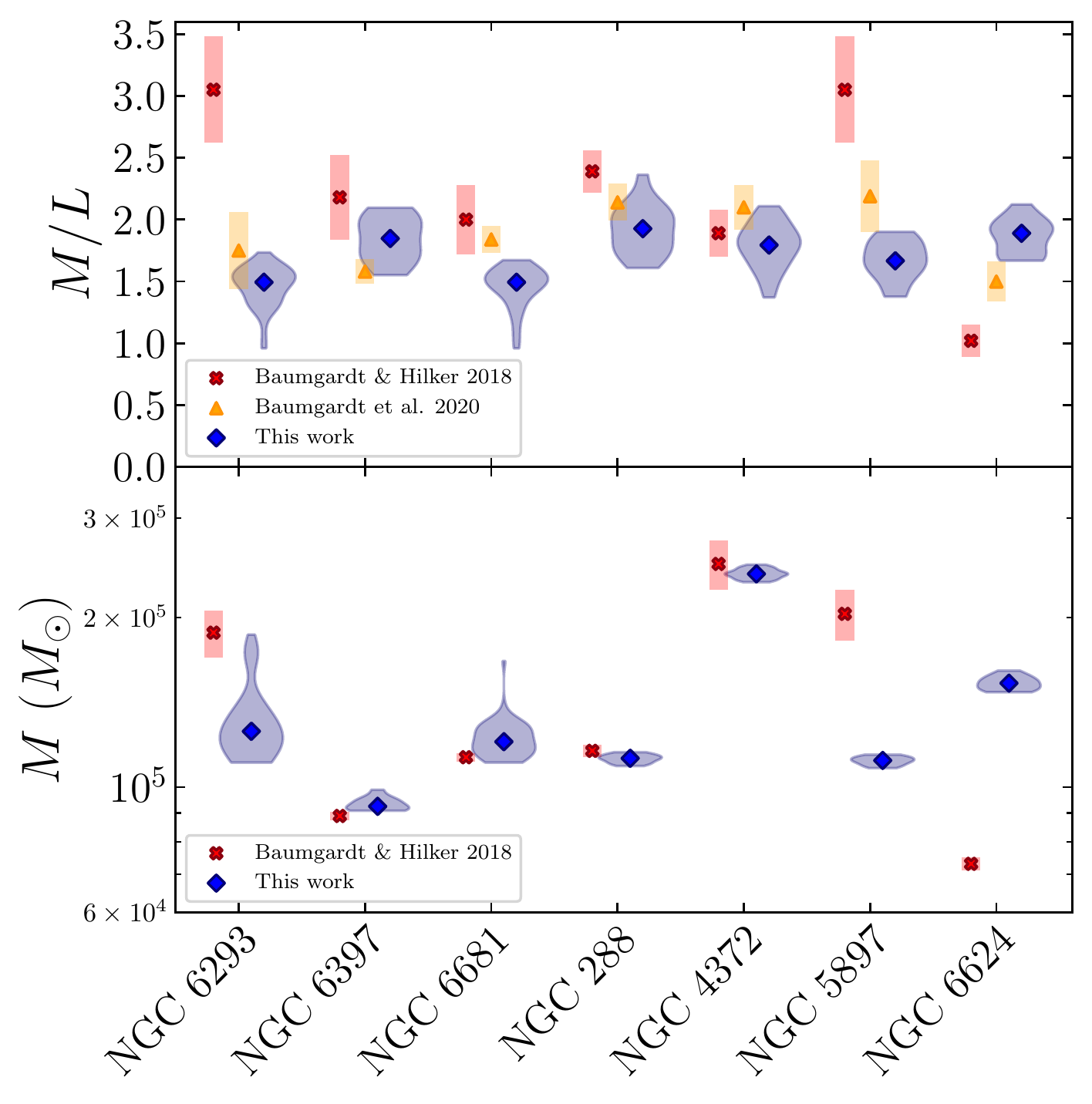}
    \caption{Violin plots of our estimates of the mass-to-light ratios (\textit{top}) and total cluster masses (\textit{bottom}) for core-collapsed clusters NGC 6293, NGC 6397, NGC 6681, and NGC 6624, and non-core-collapsed clusters NGC 288, NGC 4372, and NGC 5897.
    The widths of the ``violins'' represent the density of snapshots with a given value of $M/L$ or $M$.
    Estimates of $M/L$ and $M$ from \citet{baumgardt2018catalogue} using $N$-body simulations combined with scaling relations and updated estimates of $M/L$ from \citet{baumgardt2020absolute} using refined measurements of cluster brightnesses are also shown, where the error bars represent their $1\sigma$ uncertainties.
    }
    \label{fig:mass_plot}
\end{figure}

Though cluster brightness and extent are readily observable quantities, their precise translation to total cluster mass is complicated and generally requires dynamical modeling.
Using scaled-up versions of $N$-body simulations with $N=(1$--$2)\times10^5$, \citet{baumgardt2018catalogue} estimate the total masses and mass-to-light ratios of a number of GCs using radial velocities, including the specific clusters discussed in this paper.
They refine the mass-to-light calculations further with additional measurements of $L_V$, where they find typical values $M/L_V\sim1.8$ \citep{baumgardt2020absolute}.
We perform analogous estimates for the total cluster mass $M$ and mass-to-light ratio $M/L$ where $L$ refers to the bolometric luminosity (Figure \ref{fig:mass_plot}).

The seven clusters we examine are estimated to have present-day masses ranging between $\sim9.2\times10^4$ $M_\odot$ and $\sim1.1\times10^5$ $M_\odot$, and $M/L$ between $\sim1.5$ and $\sim1.9$ (Table \ref{tab:masses}).
Our mass estimates are very consistent with those of \citet{baumgardt2020absolute}, except in the cases of NGC 5897, where they estimate a significantly larger mass $M\sim2\times10^5$ $M_\odot$, and NGC 6624, where they estimate a significantly lower mass $M\sim7\times10^4$ $M_\odot$.
Our mass-to-light ratios are very consistent with those of \citet{baumgardt2020absolute} in all cases, and are narrowly scattered around $M/L\sim1.8$.

\begin{table*}
\caption{Masses and Mass-to-Light Ratios for Seven GCs}\label{tab:masses}
\begin{center}
\begin{tabular}{l l l l l l}
\toprule
& \multicolumn{2}{c}{Mass $M$ ($M_\odot$)} & \multicolumn{3}{c}{Mass-to-Light Ratio $M/L$ ($M_\odot/L_\odot$)} \\
\cmidrule(lr){2-3}\cmidrule(lr){4-6}
Cluster & This work & \citet{baumgardt2018catalogue} & This work & \citet{baumgardt2018catalogue} & \citet{baumgardt2020absolute} \\
\midrule
NGC 6293 & $\left(1.26^{+0.61}_{-0.15}\right)\times10^5$ & $(1.88\pm0.18)\times10^5$ & $1.49^{+0.24}_{-0.53}$ & $1.67\pm0.29$ & $1.75\pm0.31$ \\
NGC 6397 & $\left(9.25^{+0.65}_{-0.16}\right)\times10^4$ & $(8.89\pm0.16)\times10^4$ & $1.85^{+0.25}_{-0.30}$ & $2.18\pm0.34$ & $1.58\pm0.10$ \\
NGC 6681 & $\left(1.20^{+0.47}_{-0.10}\right)\times10^5$ & $(1.13\pm0.02)\times10^5$ & $1.49^{+0.18}_{-0.53}$ & $2.00\pm0.28$ & $1.84\pm0.11$ \\
NGC 288 & $\left(1.13^{+0.03}_{-0.03}\right)\times10^5$ & $(1.16\pm0.03)\times10^5$ & $1.93^{+0.44}_{-0.32}$ & $2.39\pm0.17$ & $2.14\pm0.15$ \\
NGC 4372 & $\left(2.39^{+0.09}_{-0.08}\right)\times10^5$ & $(2.49\pm0.25)\times10^5$ & $1.79^{+0.31}_{-0.42}$ & $1.89\pm0.19$ & $2.10\pm0.18$ \\
NGC 5897 & $\left(1.12^{+0.03}_{-0.03}\right)\times10^5$ & $(2.03\pm0.21)\times10^5$ & $1.67^{+0.23}_{-0.29}$ & $3.05\pm0.43$ & $2.19\pm0.29$ \\
NGC 6624 & $\left(1.53^{+0.08}_{-0.06}\right)\times10^5$ & $(7.31\pm0.20)\times10^4$ & $1.89^{+0.23}_{-0.22}$ & $1.02\pm0.13$ & $1.50\pm0.16$ \\
\bottomrule
\end{tabular}
\end{center}

Cluster mass and mass-to-light ratio for NGC 6293, NGC 6397, NGC 6681, NGC 288, NGC 4372, NGC 5897, and NGC 6624.
The uncertainty bars reported here are taken to span the entire range of values for $M$ and $M/L$ which appear in well-fitting snapshots for a given cluster.
Values of $M$ from \citet{baumgardt2018catalogue} as well as $M/L$ from \citet{baumgardt2018catalogue} and \citet{baumgardt2020absolute} for these clusters are also reproduced above.
\end{table*}

\subsection{Binary Fraction} \label{binary}

\begin{figure}
    \centering
    \includegraphics[width=0.45\textwidth]{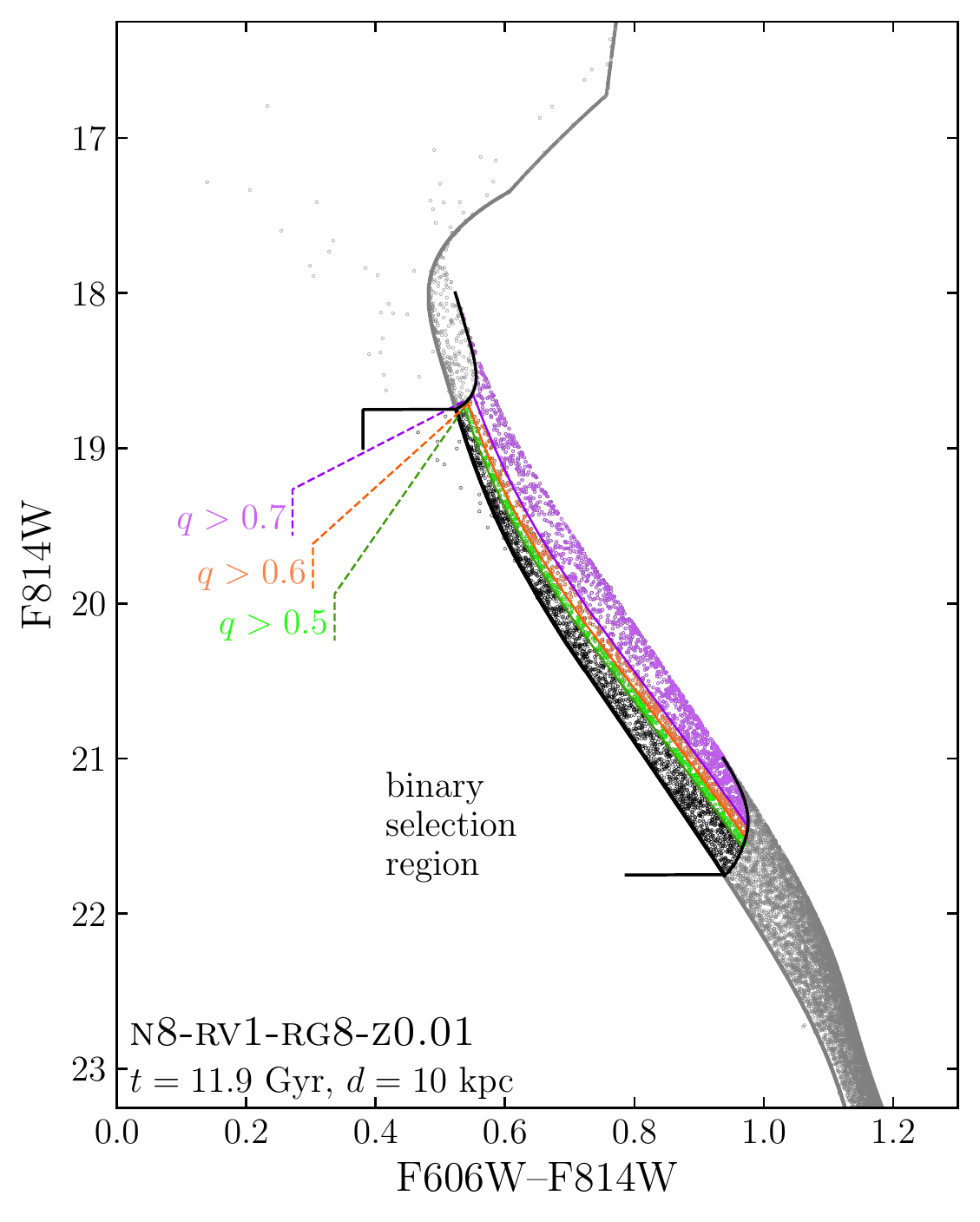}
    \caption{A synthetic color-magnitude diagram showing the cuts applied to the simulated catalog to find binaries with $q>0.5$, $q>0.6$, and $q>0.7$.
    We apply cuts to mimic the observational procedure of \citet{milone2012acs} (ACS Globular Cluster Survey) as closely as possible.}
    \label{fig:cuts_plot}
\end{figure}

Within GCs, the binary fraction is photometrically observable property which is sensitive to cluster dynamics, particularly in their cores.
The dense environments provided by GCs frequently scatter and eject binary systems, and the dynamical formation of binaries is generally thought to be the halting mechanism for collapse in core-collapsed clusters after the expulsion of their black holes \citep[e.g.,][]{chatterjee2013understanding}.
Moreover, as binary systems almost always have total fluxes equal to the sum of their component fluxes, main-sequence binary systems can be found on the color-magnitude diagram in predictably brighter sequences above the main-sequence defined by their mass ratios.
In a GC, binaries can either persist from the initial formation of the cluster or be dynamically generated over time---in the \texttt{CMC Cluster Catalog}, it is assumed in all cases that the primordial binary fraction is $5\%$ with a flat mass ratio between $q=0.1$ and $1$.
Using the ACS Globular Cluster Survey, \citet{milone2012acs} photometrically measure the binary fraction for $59$ GCs for mass ratios $q>0.5$, $0.6$, and $0.7$ individually within three radial regions $r<r_c$, $r_c<r<r_h$, and $r>r_h$, where $r_c$ and $r_h$ are given by \citet[2010 edition]{harris1996catalog} (Figure \ref{fig:cuts_plot}).
Of these, $5$ overlap with our $7$ clusters of interest (they do not report binary fractions for NGC 6293 and NGC 4372).

\begin{figure*}
    \centering
    \includegraphics[width=\textwidth]{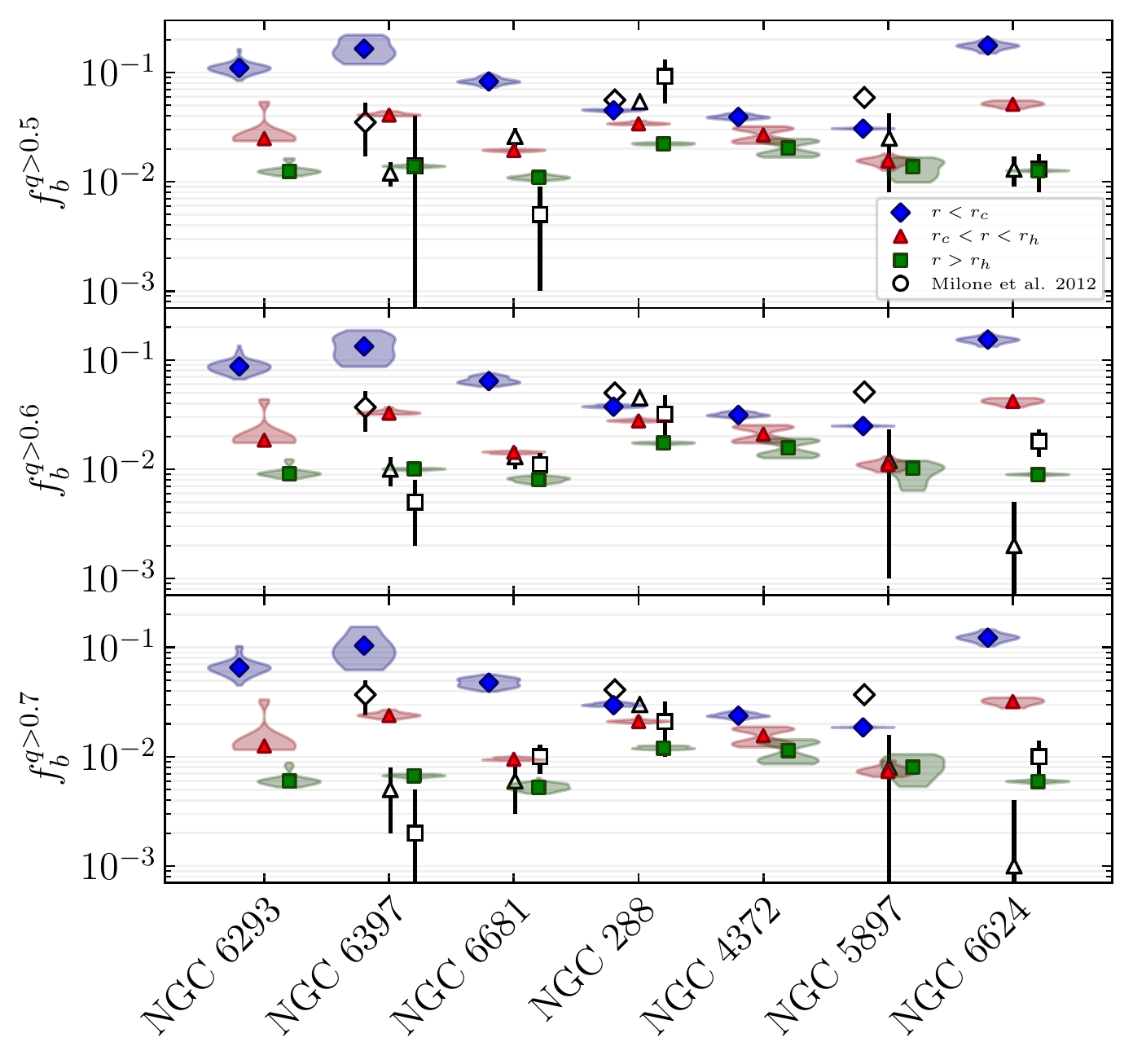}
    \caption{Binary fraction of seven GCs with mass fractions $q>0.5$ (\textit{top}), $q>0.6$ (\textit{center}), and $q>0.7$ (\textit{bottom}) with $r<r_c$ (blue), $r_c<r<r_h$ (red), and $r>r_h$ (green).
    For the model values, the horizontal width of the violin plot point refers to the density of well-fitting snapshots with that particular binary fraction value.
    Observed values given by \citet{milone2012acs} (ACS Globular Cluster Survey) are also shown.}
    \label{fig:binaries_plot}
\end{figure*}

We calculate the binary fraction subject to the same minimum mass ratios and radial ranges for all best-fitting snapshots for each of our seven clusters of interest (Figure \ref{fig:binaries_plot}).
To mimic the magnitude cuts applied by \citet{milone2012acs}, we restrict our sample to a locus on the color-magnitude diagram consistent with binaries whose primary has an F814W magnitude between $0.75$ and $3.75$ mag below the main-sequence turn-off.
On the blue edge, we enforce that included sources must have an F606W--F814W color which lies no bluer than $0.1$ mag of the main-sequence turnoff.
To calculate a binary fraction for stars above a mass ratio $q$, we use the \textsc{sse} main-sequence prescription to define a locus on the color-magnitude diagram corresponding to the sum of fluxes due to two main-sequence stars with mass ratio $q$.
Sources on the red side of the locus are then considered binaries.
The binary fraction is then calculated by dividing the weight of sources identified as binaries by the total weight of all sources in the magnitude and radial range, with weights defined by Equation \ref{eqn:p}.
Notably, while \citet{milone2012acs} additionally apply a general inner radius cut for a number of clusters, out of our $7$ clusters of interest only NGC 6681 and NGC 6624 are affected by such a cut, and in particular only the $r_c<r<r_h$ annulus (they do not report binary fractions for $r<r_c$).
However, in order to keep applied cuts relatively consistent between the clusters, we omit this cut for these particular GCs.

We find reasonable consistency between the model binary fractions and the data in most cases.
One notable exception is the binary fraction within the inner spatial bins of NGC 6397, where our models appear to predict binary fractions of up to $\sim3$ times the observed value--even here, the data and models come back into agreement in the outermost bin.
Another is in the $r_c<r<r_h$ bin of NGC 6624, where \citet{milone2012acs} report very low binary fractions which are curiously lower than the binary fraction of the outskirts of the cluster ($r>r_h$).
However, in general, measurement of the binary fraction within core-collapsed clusters is relatively difficult for a number of reasons.
First of all, formally, core-collapsed clusters do not have a well-defined core radius, and observational definitions of the core radius tend to be quite small (e.g., $r_c\sim0.03$ pc for NGC 6397), with conversions from angular units to physical distances being very sensitive to heliocentric distance.
As binaries are dynamically generated and then burned in large numbers in the cores of core-collapsed clusters, binary fraction is expected to vary substantially with respect to distance from the GC center.
Moreover, in the central regions of such clusters, stellar density is extremely large, implying relatively low completeness.

Overall, we note acceptable agreement between our predictions and observations for non-core-collapsed clusters as well as the outer regions of core-collapsed clusters.
This provides a degree of reassurance that \texttt{CMC} can sensibly replicate the binary populations of realistic GCs, although the fixed assumed binary fraction of $5\%$ in a flat mass ratio distribution combined with a complex initial to final binary fraction mapping complicates the picture somewhat.

\subsection{Mass Segregation} \label{massseg}

\begin{figure}
    \centering
    \includegraphics[width=0.45\textwidth]{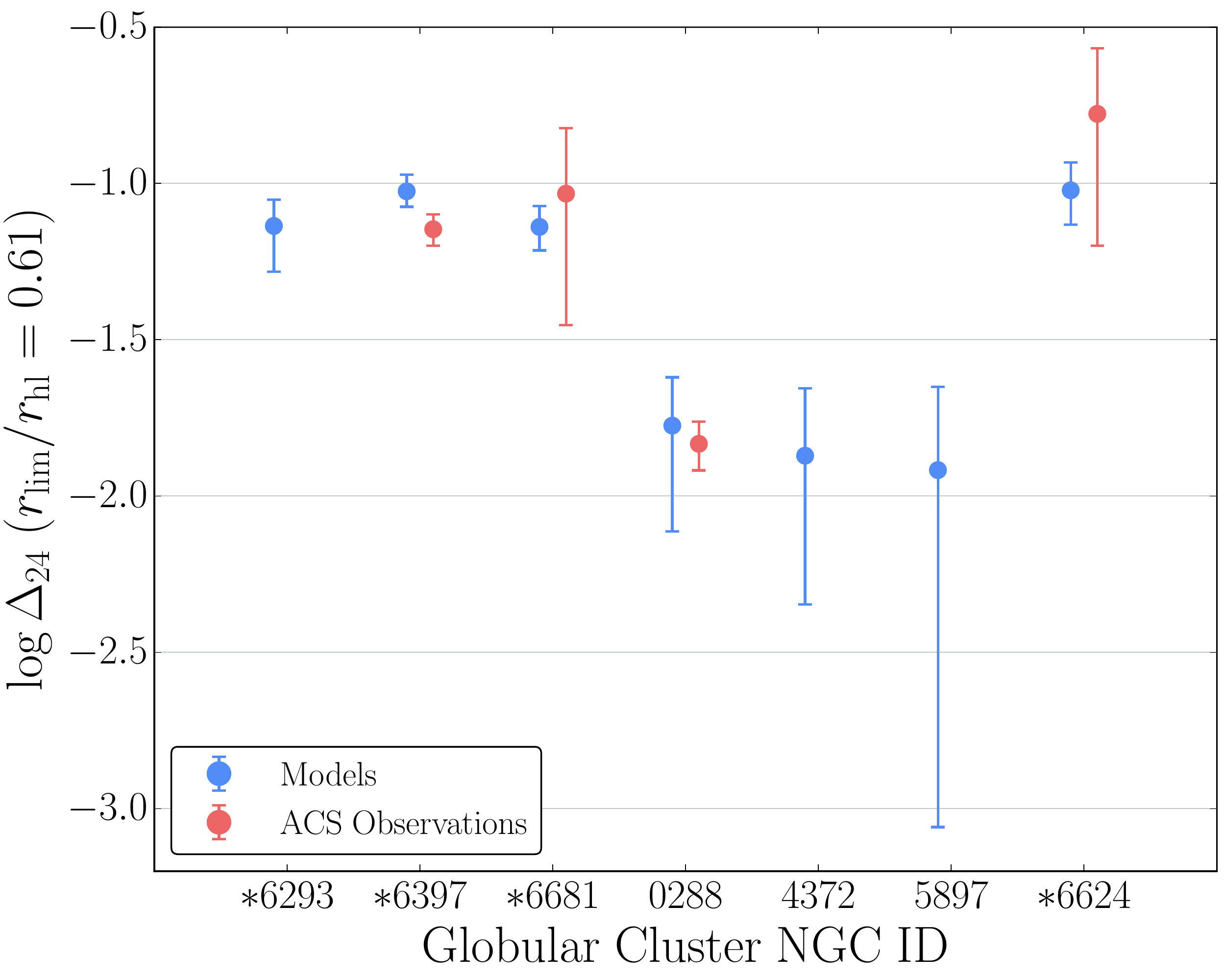}
    \caption{The mass segregation metric $\Delta_{24}$ calculated for both well-fitting snapshots to our $7$ GCs of interest and also directly from ACS Globular Cluster Survey data where available.
In both cases, the plotted uncertainties correspond to a 95\% confidence interval, with the uncertainty in the model values taking into account variations due to different two-dimensional projections as estimated by $10$ different realizations per snapshot.
General agreement between the simulated and observed values is apparent.
}
    \label{fig:dr24_comparison_rhl61}
\end{figure}

Over thermodynamically long timescales, massive stars are expected to sink to the center of the cluster through dynamical friction.
It can be shown that a population of stars with mass $m$ will segregate within a cluster on a timescale $t_{m,\mathrm{MS}}\sim\left(\left\langle m\right\rangle/m\right)t_{\mathrm{rlx}}$ where $\left\langle m\right\rangle$ is the mean mass of the cluster \citep{portegies2010young}.
As $t_{\mathrm{rlx}}\sim\mathrm{few}\times10^9$ years for typical GCs, mass segregation in GCs is readily identified as a preferential clustering of massive stars closer to the center of the cluster.
Moreover, as ``dark'' objects such as black holes and other stellar remnants also participate in mass segregation, they may influence observed metrics of mass segregation in nontrivial ways.

We examine here the ability of the SBP and VDP alone to predict the degree of mass segregation within a cluster.
In particular, in accordance with \citet{weatherford2019dynamical}, we define Population II stars as the ``high-mass'' population with $L_{\mathrm{MSTO}}/5<L<L_{\mathrm{MSTO}}$, and Population IV stars as the ``low-mass'' population with $L_{\mathrm{MSTO}}/125<L<L_{\mathrm{MSTO}}/25$, where $L_{\mathrm{MSTO}}$ is the luminosity of the main-sequence turnoff.
We then parameterize the mass segregation for each cluster $\Delta_{24}$, defined to be the difference between the median projected radial distance of Population II stars and Population IV stars, normalized by the half-light radius of the cluster.

Note that $\Delta_{24}$ is straightforwardly calculated both for simulations and observed clusters.
In particular, using the ACS Globular Cluster Survey, we reproduce this calculation for the $4$ of our $7$ clusters of interest which have suitable observations, following the procedure of \citet{weatherford2018predicting} and \citet{weatherford2019dynamical}.
We take into account incompleteness in the observed catalogs estimated from artificial star tests.
We then compare these to the distribution of $\Delta_{24}$ within well-fitting snapshots for our $7$ clusters (Figure \ref{fig:dr24_comparison_rhl61}).
Within the simulated data, we mimic the limited field of view of the data by restricting to stars within a projected radius of 61\% of the half-light radius (which is the highest value of the $4$ clusters which can be accommodated by the data).
In cases where comparison is possible, we find very strong agreement between the simulated and observed $\Delta_{24}$ except in NGC 6397, where the simulations slightly overestimate the degree of mass segregation.

\subsection{Black Holes} \label{bh}

\begin{table*}
\caption{Predicted Black Hole Counts for Seven GCs}\label{tab:bh}
\begin{tabular}{l l l l l l l l l l l}
\toprule
& \multicolumn{5}{c}{This work} & \multicolumn{5}{c}{\citet{weatherford2019dynamical}} \\
\cmidrule(lr){2-6}\cmidrule(lr){7-11}
Cluster & Min. & $-1\sigma$ & Median & $+1\sigma$ & Max. & $-2\sigma$ & $-1\sigma$ & Median & $+1\sigma$ & $+2\sigma$ \\
\midrule
NGC 6293$^c$ & $0.00$ & $0.00$ & $0.00$ & $0.00$ & $16.00$ & --- & --- & --- & --- & --- \\
NGC 6397$^c$ & $1.00$ & $1.00$ & $1.00$ & $1.00$ & $1.00$ & $0.00$ & $0.00$ & $0.61$ & $1.80$ & $4.06$ \\
NGC 6681$^c$ & $0.00$ & $0.00$ & $0.00$ & $0.00$ & $7.00$ & $0.00$ & $1.21$ & $5.02$ & $10.10$ & $16.30$ \\
NGC 288 & $48.00$ & $54.40$ & $70.00$ & $79.20$ & $88.00$ & $2.24$ & $9.93$ & $18.2$ & $26.6$ & $46.9$ \\
NGC 4372 & $93.00$ & $106.44$ & $217.00$ & $347.68$ & $375.00$ & --- & --- & --- & --- & --- \\
NGC 5897 & $44.00$ & $44.24$ & $56.00$ & $64.76$ & $67.00$ & --- & --- & --- & --- & --- \\
NGC 6624$^c$ & $2.00$ & $2.00$ & $6.00$ & $6.00$ & $9.00$ & $0.70$ & $19.60$ & $23.20$ & $26.80$ & $31.1$ \\
\bottomrule
\end{tabular}

The number of BHs in well-fitting snapshots for $7$ Milky Way GCs.
For each cluster, the median, maximum and minimum, $16$th and $84$th percentiles are reported.
Core-collapsed clusters are identified using a subscript $c$.
For reference, we have also included the number of black holes as estimated by \citet{weatherford2019dynamical} using the observed mass segregation.
\end{table*}

While black holes (BHs) in GCs are very difficult to detect directly, their presence and number can be indirectly inferred by examining their effect on a GC's dynamical state.
For example, by considering a grid of \texttt{CMC} models finely gridded over initial virial radius, \citet{kremer2019initial} demonstrate the importance of BHs in the halting of core collapse in GCs NGC 3201, NGC 6656 (M22), and NGC 6254 (M10), and note their likely absence in NGC 6752, which is core-collapsed.
These BH populations are, in turn, intimately related to the cluster's initial size.
Along a similar vein, trends in the core radius with age for massive clusters in Milky Way satellite galaxies in the seminal Mackey \& Gilmore catalogs \citep{lmc2003surface,smc2003surface,fornax2003surface} have been interpreted as evidence for the role of BHs in the clusters' bulk evolution \citep[see, e.g.,][which reproduces these observed trends in $N\sim10^5$-body simulations]{mackey2008black}.

Evidence of the effect of BH populations on the structure of a GC have also been evident in the spatial distribution of stars in different mass ranges.
In particular, using \texttt{CMC} models, \citet{weatherford2019dynamical} constrain the number of BHs in GCs by taking advantage of an anticorrelation between the extent of mass segregation in a cluster and its BH population, a trend quantifiable in \texttt{CMC} models.
Intuitively, this anticorrelation arises from the rapid segregation of a GC's black hole population followed by dynamical heating of the massive star population and their typical distances from the cluster center.

For a similar reason, the presence of a BH population in a GC halts core collapse---large BH populations transfer significant energy to their host cluster's stellar population through binary-mediated dynamics, preventing core collapse of the bulk stellar population \citep[e.g.,][]{kremer2020iau}.
This manifests both in a flattened core surface brightness as well as a heightened dynamical temperature in the core of the cluster.
This motivates the use of observed SBPs and VDPs to constrain the BH populations of GCs, which can in turn be done by analyzing the simulated stellar populations of their best-fitting \texttt{CMC} models.

For each of the $7$ GCs of interest, we calculate the median, $18$th ($-1\sigma$) and $84$th ($+1\sigma$) percentiles, and minimum and maximum number of BH in well-fitting snapshots (Table \ref{tab:bh}).
As expected, all three of the core-collapse clusters examined have fully single-digit BH counts.
Of their sample of $50$ GCs, $4$ of the GCs for which \citet{weatherford2019dynamical} have estimated BH counts coincide with our $7$: NGC 6397, NGC 6681, NGC 288, and NGC 6624.
Reassuringly, our BH predictions are consistent with theirs when the metric parameterizing mass segregation is consistent with the definition in Section \ref{massseg}.
Moreover, both \citet{weatherford2019dynamical} and our work broadly reflect the tendency of core-collapsed clusters to have fewer BHs, reiterating the story that black holes provide the dominant mechanism for halted core collapse in the majority of GCs today. Though both this work and that of \citet{weatherford2019dynamical} calibrate $N_{\mathrm{BH}}$ to the same grid of models, we obtain constraints from two distinct observables (the SBP and VDP versus the degree of mass segregation in the cluster).
This indicates at least concordance with the idea that both the (suppressed) degree of mass segregation and cluster dynamics are broadly driven by a single BH population at the center of a GC (or the lack thereof).
Nevertheless, given a lack of direct observations of $N_{\rm BH}$, the actual size of this population remains highly uncertain.

\section{Conclusion}

The approach to GC modeling enabled by \texttt{CMC} provides a balanced approach to running accurate, long-timescale simulations of realistically large GCs in practical runtimes without reliance on scaling relations for deducing cluster parameters.
This opens the door to holistic, direct comparisons of observations to extensive model grids over realistic GC parameter spaces.
Accordingly, we present a scheme for identifying well-fitting simulation snapshots from the \texttt{CMC Cluster Catalog}.
Out of $59$ Milky Way GCs, we find that the \texttt{CMC Cluster Catalog} provides good fits to $26$ GCs as is.
As illustrative examples, we focus specifically on six of these well-fit clusters in our database.
In order to demonstrate that the number of good fits can be extended straightforwardly with the addition of new \texttt{CMC} models, we detail a procedure for augmenting the model grid to fit a seventh GC, NGC 6624, which is not well-fit by any snapshot on the original \texttt{CMC Cluster Catalog}.
We examine the clusters' predicted masses, mass-to-light ratios, binary fractions, and black hole counts, finding reasonable consistency with previous works and observations in most cases.
The predicted numbers of of cataclysmic variables, low-mass X-ray binaries, and millisecond pulsars are also reported when analogous observations exist, with consistency in all cases except possibly in the case of NGC 6624.

Motivated by the desire to extend the utility of this method to a wider range of clusters as well as the precision of the comparison, we suggest a number of potential refinements to this procedure, namely: (1) extension of the grid both to parameters within the current parameter range (to increase the parameter space grid resolution) and outside (to extend the grid to fit GCs which are not represented in the current model grid), (2) variation of additional parameters such as the binarity or initial mass function in order to better capture the full diversity of possible GC evolution histories, (3) including the observed mass function slope in constraints on the GC as a further axis of comparison \citep[such observations are already available for a number of Milky Ways GCs, e.g.,][]{sollima2017global}, and (4) proactively leveraging observed stellar counts for populations such as CVs, XRBs, and blue stragglers as additional constraints in matching models.

We make available a set of functions for analyzing GC models generated using \texttt{CMC}, including the already publicly available \texttt{CMC Cluster Catalog}\footnote{https://cmc.ciera.northwestern.edu/home/}, as well as files containing the model SBPs, VDPs, and other parameters for the snapshots considered in this work.

\acknowledgments
We thank L.~Clifton Johnson for invaluable discussion and advice\response{, and the anonymous referee for their useful suggestions}.
This work was supported by NSF grant AST-1716762 and through the computational resources and staff contributions provided for the Quest high-performance computing facility at Northwestern University.
NZR acknowledges support from the Illinois Space Grant Consortium and the Dominic Orr Graduate Fellowship.
KK is supported by an NSF Astronomy and Astrophysics Postdoctoral Fellowship under award AST-2001751.
\response{NCW acknowledges support from the CIERA Riedel Graduate Fellowship at Northwestern University as well as the NSF GK-12 Fellowship Program under Grant DGE-0948017.
SC acknowledges support of the Department of Atomic Energy, Government of India, under  project no.\ 12-R\&D-TFR-5.02-0200.}

\software{Astropy \citep{astropy_collab}, IPython \citep{ipython}, Matplotlib \citep{hunter2007matplotlib}, NumPy \citep{numpy}, SciPy \citep{jones-scipy}, Pandas \citep{mckinney-proc-scipy-2010}, \texttt{Cluster Monte Carlo} \citep{pattabiraman2013parallel}, \response{\texttt{cmctoolkit}} \citep{codetag}, \textsc{cosmic} \citep{breivik2020cosmic}, \textsc{fewbody} \citep{fregeau2004stellar}}

%\vspace{3cm} % If this is removed, there is a bug where the acknowledgments overlap with the text

\bibliographystyle{aasjournal.bst}
\bibliography{matching}

\begin{thebibliography}{}
\expandafter\ifx\csname natexlab\endcsname\relax\def\natexlab#1{#1}\fi
\providecommand{\url}[1]{\href{#1}{#1}}

\bibitem[{Aarseth \& Heggie(1998)}]{aarseth1998basic}
Aarseth, S., \& Heggie, D. 1998, Monthly Notices of the Royal Astronomical
  Society, 297, 794

\bibitem[{Allard {et~al.}(1994)Allard, Hauschildt, Miller, \&
  Tennyson}]{allard1994influence}
Allard, F., Hauschildt, P., Miller, S., \& Tennyson, J. 1994, The Astrophysical
  Journal, 426, L39

\bibitem[{Bae {et~al.}(2014)Bae, Kim, \& Lee}]{bae2014compact}
Bae, Y.-B., Kim, C., \& Lee, H.~M. 2014, Monthly Notices of the Royal
  Astronomical Society, 440, 2714

\bibitem[{Baraffe {et~al.}(1995)Baraffe, Chabrier, Allard, \&
  Hauschildt}]{baraffe1995new}
Baraffe, I., Chabrier, G., Allard, F., \& Hauschildt, P. 1995, The
  Astrophysical Journal, 446, L35

\bibitem[{Baumgardt(2001)}]{baumgardt2001scaling}
Baumgardt, H. 2001, Monthly Notices of the Royal Astronomical Society, 325,
  1323

\bibitem[{Baumgardt \& Hilker(2018)}]{baumgardt2018catalogue}
Baumgardt, H., \& Hilker, M. 2018, Monthly Notices of the Royal Astronomical
  Society, 478, 1520

\bibitem[{Baumgardt {et~al.}(2019{\natexlab{a}})Baumgardt, Hilker, Sollima, \&
  Bellini}]{baumgardt2018mean}
Baumgardt, H., Hilker, M., Sollima, A., \& Bellini, A. 2019{\natexlab{a}},
  Monthly Notices of the Royal Astronomical Society, 482, 5138

\bibitem[{Baumgardt {et~al.}(2020)Baumgardt, Sollima, \&
  Hilker}]{baumgardt2020absolute}
Baumgardt, H., Sollima, A., \& Hilker, M. 2020, arXiv preprint arXiv:2009.09611

\bibitem[{Baumgardt {et~al.}(2019{\natexlab{b}})Baumgardt, He, Sweet,
  Drinkwater, Sollima, Hurley, Usher, Kamann, Dalgleish, Dreizler,
  {et~al.}}]{baumgardt2019no}
Baumgardt, H., He, C., Sweet, S.~M., {et~al.} 2019{\natexlab{b}}, Monthly
  Notices of the Royal Astronomical Society, 488, 5340

\bibitem[{Bellazzini {et~al.}(2001)Bellazzini, Pecci, Ferraro, Galleti,
  Catelan, \& Landsman}]{bellazzini2001age}
Bellazzini, M., Pecci, F.~F., Ferraro, F.~R., {et~al.} 2001, The Astronomical
  Journal, 122, 2569

\bibitem[{Biggs {et~al.}(1994)Biggs, Bailes, Lyne, Goss, \&
  Fruchter}]{biggs1994two}
Biggs, J., Bailes, M., Lyne, A., Goss, W., \& Fruchter, A. 1994, Monthly
  Notices of the Royal Astronomical Society, 267, 125

\bibitem[{{Breivik} {et~al.}(2020){Breivik}, {Coughlin}, {Zevin}, {Rodriguez},
  {Kremer}, {Ye}, {Andrews}, {Kurkowski}, {Digman}, {Larson}, \&
  {Rasio}}]{Breivik2020}
{Breivik}, K., {Coughlin}, S., {Zevin}, M., {et~al.} 2020, \apj, 898, 71

\bibitem[{Breivik {et~al.}(2020)Breivik, Coughlin, Zevin, Rodriguez, Kremer,
  Claire, Andrews, Kurkowski, Digman, Larson, {et~al.}}]{breivik2020cosmic}
Breivik, K., Coughlin, S., Zevin, M., {et~al.} 2020, The Astrophysical Journal,
  898, 71

\bibitem[{Cardelli {et~al.}(1989)Cardelli, Clayton, \&
  Mathis}]{cardelli1989relationship}
Cardelli, J.~A., Clayton, G.~C., \& Mathis, J.~S. 1989, The Astrophysical
  Journal, 345, 245

\bibitem[{Casagrande \& VandenBerg(2014)}]{casagrande2014synthetic}
Casagrande, L., \& VandenBerg, D.~A. 2014, Monthly Notices of the Royal
  Astronomical Society, 444, 392

\bibitem[{Chatterjee {et~al.}(2010)Chatterjee, Fregeau, Umbreit, \&
  Rasio}]{chatterjee2010monte}
Chatterjee, S., Fregeau, J.~M., Umbreit, S., \& Rasio, F.~A. 2010, The
  Astrophysical Journal, 719, 915

\bibitem[{Chatterjee {et~al.}(2013)Chatterjee, Umbreit, Fregeau, \&
  Rasio}]{chatterjee2013understanding}
Chatterjee, S., Umbreit, S., Fregeau, J.~M., \& Rasio, F.~A. 2013, Monthly
  Notices of the Royal Astronomical Society, 429, 2881

\bibitem[{Chen \& Chen(2010)}]{chen2010morphological}
Chen, C., \& Chen, W. 2010, The Astrophysical Journal, 721, 1790

\bibitem[{Chen \& Han(2004)}]{chen2004effects}
Chen, X., \& Han, Z. 2004, Monthly Notices of the Royal Astronomical Society,
  355, 1182

\bibitem[{Cohn {et~al.}(2010)Cohn, Lugger, Couch, Anderson, Cool, Van~den Berg,
  Bogdanov, Heinke, \& Grindlay}]{cohn2010identification}
Cohn, H.~N., Lugger, P.~M., Couch, S.~M., {et~al.} 2010, The Astrophysical
  Journal, 722, 20

\bibitem[{Cool {et~al.}(1995)Cool, Grindlay, Cohn, Lugger, \&
  Slavin}]{cool1995discovery}
Cool, A.~M., Grindlay, J.~E., Cohn, H.~N., Lugger, P.~M., \& Slavin, S.~D.
  1995, The Astrophysical Journal, 439, 695

\bibitem[{Cool {et~al.}(1996)Cool, Piotto, \& King}]{cool1996main}
Cool, A.~M., Piotto, G., \& King, I.~R. 1996, The Astrophysical Journal, 468,
  655

\bibitem[{Dieball {et~al.}(2017)Dieball, Rasekh, Knigge, Shara, \&
  Zurek}]{dieball2017far}
Dieball, A., Rasekh, A., Knigge, C., Shara, M., \& Zurek, D. 2017, Monthly
  Notices of the Royal Astronomical Society, 469, 267

\bibitem[{Fabbiano(2006)}]{fabbiano2006populations}
Fabbiano, G. 2006, Annual Review of Astronomy and Astrophysics, 44

\bibitem[{Ferraro {et~al.}(2012)Ferraro, Lanzoni, Dalessandro, Beccari,
  Pasquato, Miocchi, Rood, Sigurdsson, Sills, Vesperini,
  {et~al.}}]{ferraro2012dynamical}
Ferraro, F., Lanzoni, B., Dalessandro, E., {et~al.} 2012, Nature, 492, 393

\bibitem[{Ferraro {et~al.}(2018{\natexlab{a}})Ferraro, Mucciarelli, Lanzoni,
  Pallanca, Lapenna, Origlia, Dalessandro, Valenti, Beccari, Bellazzini,
  {et~al.}}]{ferraro2018mikis}
Ferraro, F., Mucciarelli, A., Lanzoni, B., {et~al.} 2018{\natexlab{a}}, The
  Astrophysical Journal, 860, 50

\bibitem[{Ferraro {et~al.}(2018{\natexlab{b}})Ferraro, Lanzoni, Raso,
  Nardiello, Dalessandro, Vesperini, Piotto, Pallanca, Beccari, Bellini,
  {et~al.}}]{ferraro2018hubble}
Ferraro, F., Lanzoni, B., Raso, S., {et~al.} 2018{\natexlab{b}}, The
  Astrophysical Journal, 860, 36

\bibitem[{Fregeau {et~al.}(2004)Fregeau, Cheung, Portegies~Zwart, \&
  Rasio}]{fregeau2004stellar}
Fregeau, J.~M., Cheung, P., Portegies~Zwart, S., \& Rasio, F.~A. 2004, Monthly
  Notices of the Royal Astronomical Society, 352, 1

\bibitem[{Fregeau {et~al.}(2003)Fregeau, G{\"u}rkan, Joshi, \&
  Rasio}]{fregeau2003monte}
Fregeau, J.~M., G{\"u}rkan, M., Joshi, K., \& Rasio, F. 2003, The Astrophysical
  Journal, 593, 772

\bibitem[{Fregeau \& Rasio(2007)}]{fregeau2007monte}
Fregeau, J.~M., \& Rasio, F.~A. 2007, The Astrophysical Journal, 658, 1047

\bibitem[{Gieles {et~al.}(2018)Gieles, Balbinot, ISM~Yaaqib,
  H{\'e}nault-Brunet, Zocchi, Peuten, \& Jonker}]{gieles2018mass}
Gieles, M., Balbinot, E., ISM~Yaaqib, R., {et~al.} 2018, Monthly Notices of the
  Royal Astronomical Society, 473, 4832

\bibitem[{Gieles {et~al.}(2021)Gieles, Erkal, Antonini, Balbinot, \&
  Pe{\~n}arrubia}]{gieles2021supra}
Gieles, M., Erkal, D., Antonini, F., Balbinot, E., \& Pe{\~n}arrubia, J. 2021,
  arXiv preprint arXiv:2102.11348

\bibitem[{Giersz(1998)}]{giersz1998monte}
Giersz, M. 1998, Monthly Notices of the Royal Astronomical Society, 298, 1239

\bibitem[{Giersz \& Heggie(2009)}]{giersz2009monte}
Giersz, M., \& Heggie, D.~C. 2009, Monthly Notices of the Royal Astronomical
  Society, 395, 1173

\bibitem[{Giersz \& Heggie(2011)}]{giersz2011monte}
---. 2011, Monthly Notices of the Royal Astronomical Society, 410, 2698

\bibitem[{Giersz {et~al.}(2013)Giersz, Heggie, Hurley, \&
  Hypki}]{giersz2013mocca}
Giersz, M., Heggie, D.~C., Hurley, J.~R., \& Hypki, A. 2013, Monthly Notices of
  the Royal Astronomical Society, 431, 2184

\bibitem[{Giersz {et~al.}(2015)Giersz, Leigh, Hypki, L{\"u}tzgendorf, \&
  Askar}]{giersz2015mocca}
Giersz, M., Leigh, N., Hypki, A., L{\"u}tzgendorf, N., \& Askar, A. 2015,
  Monthly Notices of the Royal Astronomical Society, 454, 3150

\bibitem[{Glebbeek {et~al.}(2008)Glebbeek, Pols, \&
  Hurley}]{glebbeek2008evolution}
Glebbeek, E., Pols, O.~R., \& Hurley, J.~R. 2008, Astronomy \& Astrophysics,
  488, 1007

\bibitem[{Green \& Norris(1990)}]{green1990population}
Green, E., \& Norris, J.~E. 1990, The Astrophysical Journal, 353, L17

\bibitem[{Grindlay {et~al.}(2001)Grindlay, Heinke, Edmonds, Murray, \&
  Cool}]{grindlay2001chandra}
Grindlay, J., Heinke, C., Edmonds, P., Murray, S., \& Cool, A. 2001, The
  Astrophysical Journal Letters, 563, L53

\bibitem[{Han {et~al.}(2017)Han, Chun, Choudhury, Chiang, Lee, \&
  Sohn}]{han2017spatial}
Han, M., Chun, S.-H., Choudhury, S., {et~al.} 2017, Journal of Astronomy and
  Space Sciences, 34, 83

\bibitem[{Hansen {et~al.}(2007)Hansen, Anderson, Brewer, Dotter, Fahlman,
  Hurley, Kalirai, King, Reitzel, Richer, {et~al.}}]{hansen2007white}
Hansen, B.~M., Anderson, J., Brewer, J., {et~al.} 2007, The Astrophysical
  Journal, 671, 380

\bibitem[{Harris(1996)}]{harris1996catalog}
Harris, W.~E. 1996, The Astronomical Journal, 112, 1487

\bibitem[{Heggie \& Giersz(2008)}]{heggie2008monte}
Heggie, D.~C., \& Giersz, M. 2008, Monthly Notices of the Royal Astronomical
  Society, 389, 1858

\bibitem[{Heggie \& Giersz(2014)}]{heggie2014mocca}
---. 2014, Monthly Notices of the Royal Astronomical Society, 439, 2459

\bibitem[{H{\'e}non(1971)}]{henon1971monte}
H{\'e}non, M. 1971, in International Astronomical Union Colloquium, Vol.~10,
  Cambridge University Press, 151--167

\bibitem[{Hui {et~al.}(2010)Hui, Cheng, \& Taam}]{hui2010dynamical}
Hui, C., Cheng, K., \& Taam, R.~E. 2010, The Astrophysical Journal, 714, 1149

\bibitem[{Hunter(2007)}]{hunter2007matplotlib}
Hunter, J.~D. 2007, Computing In Science \& Engineering, 9, 90

\bibitem[{Hurley {et~al.}(2000)Hurley, Pols, \& Tout}]{hurley2000comprehensive}
Hurley, J.~R., Pols, O.~R., \& Tout, C.~A. 2000, Monthly Notices of the Royal
  Astronomical Society, 315, 543

\bibitem[{Hurley {et~al.}(2002)Hurley, Tout, \& Pols}]{hurley2002evolution}
Hurley, J.~R., Tout, C.~A., \& Pols, O.~R. 2002, Monthly Notices of the Royal
  Astronomical Society, 329, 897

\bibitem[{Hut \& Heggie(2003)}]{hut2003gravitational}
Hut, P., \& Heggie, D. 2003, The Gravitational Million-Body Problem,  Cambridge
  Univ. Press

\bibitem[{Hut {et~al.}(1992)Hut, McMillan, Goodman, Mateo, Phinney, Pryor,
  Richer, Verbunt, \& Weinberg}]{hut1992binaries}
Hut, P., McMillan, S., Goodman, J., {et~al.} 1992, Publications of the
  Astronomical Society of the Pacific, 104, 981

\bibitem[{Ivanova {et~al.}(2010)Ivanova, Chaichenets, Fregeau, Heinke,
  Lombardi~Jr, \& Woods}]{ivanova2010formation}
Ivanova, N., Chaichenets, S., Fregeau, J., {et~al.} 2010, The Astrophysical
  Journal, 717, 948

\bibitem[{Ivanova {et~al.}(2006)Ivanova, Heinke, Rasio, Taam, Belczynski, \&
  Fregeau}]{ivanova2006formation}
Ivanova, N., Heinke, C., Rasio, F., {et~al.} 2006, Monthly Notices of the Royal
  Astronomical Society, 372, 1043

\bibitem[{Jones {et~al.}(2001)Jones, Oliphant, Peterson,
  {et~al.}}]{jones-scipy}
Jones, E., Oliphant, T., Peterson, P., {et~al.} 2001, {SciPy}: Open source
  scientific tools for {Python}, , , [Online; accessed <today>].
\newblock \url{http://www.scipy.org/}

\bibitem[{Jonker {et~al.}(2003)Jonker, M{\'e}ndez, Nelemans, Wijnands, \& Van
  Der~Klis}]{jonker2003chandra}
Jonker, P., M{\'e}ndez, M., Nelemans, G., Wijnands, R., \& Van Der~Klis, M.
  2003, Monthly Notices of the Royal Astronomical Society, 341, 823

\bibitem[{Joshi {et~al.}(2001)Joshi, Nave, \& Rasio}]{joshi2001monte}
Joshi, K.~J., Nave, C.~P., \& Rasio, F.~A. 2001, The Astrophysical Journal,
  550, 691

\bibitem[{Joshi {et~al.}(2000)Joshi, Rasio, \& Zwart}]{joshi2000monte}
Joshi, K.~J., Rasio, F.~A., \& Zwart, S.~P. 2000, The Astrophysical Journal,
  540, 969

\bibitem[{Kacharov {et~al.}(2014)Kacharov, Bianchini, Koch, Frank, Martin,
  van~de Ven, Puzia, McDonald, Johnson, \& Zijlstra}]{kacharov2014study}
Kacharov, N., Bianchini, P., Koch, A., {et~al.} 2014, Astronomy \&
  Astrophysics, 567, A69

\bibitem[{Kaluzny \& Krzeminski(1993)}]{kaluzny1993contact}
Kaluzny, J., \& Krzeminski, W. 1993, Monthly Notices of the Royal Astronomical
  Society, 264, 785

\bibitem[{Kamann {et~al.}(2017)Kamann, Husser, Dreizler, Emsellem, Weilbacher,
  Martens, Bacon, den Brok, Giesers, Krajnovi{\'c},
  {et~al.}}]{kamann2017stellar}
Kamann, S., Husser, T.-O., Dreizler, S., {et~al.} 2017, Monthly Notices of the
  Royal Astronomical Society, 473, 5591

\bibitem[{Knigge(2012)}]{knigge2012cvs}
Knigge, C. 2012, Mem. Soc. Astron. Ital., 83

\bibitem[{Kong {et~al.}(2006)Kong, Bassa, Pooley, Lewin, Homer, Verbunt,
  Anderson, \& Margon}]{kong2006chandra}
Kong, A.~K., Bassa, C., Pooley, D., {et~al.} 2006, The Astrophysical Journal,
  647, 1065

\bibitem[{Kremer {et~al.}(2018)Kremer, Chatterjee, Breivik, Rodriguez, Larson,
  \& Rasio}]{kremer2018lisa}
Kremer, K., Chatterjee, S., Breivik, K., {et~al.} 2018, Physical review
  letters, 120, 191103

\bibitem[{{Kremer} {et~al.}(2018){Kremer}, {Chatterjee}, {Rodriguez}, \&
  {Rasio}}]{kremer2018xrb}
{Kremer}, K., {Chatterjee}, S., {Rodriguez}, C.~L., \& {Rasio}, F.~A. 2018,
  \apj, 852, 29

\bibitem[{Kremer {et~al.}(2019)Kremer, Chatterjee, Ye, Rodriguez, \&
  Rasio}]{kremer2019initial}
Kremer, K., Chatterjee, S., Ye, C.~S., Rodriguez, C.~L., \& Rasio, F.~A. 2019,
  The Astrophysical Journal, 871, 38

\bibitem[{Kremer {et~al.}(2018)Kremer, Ye, Chatterjee, Rodriguez, \&
  Rasio}]{kremer2018black}
Kremer, K., Ye, C.~S., Chatterjee, S., Rodriguez, C.~L., \& Rasio, F.~A. 2018,
  The Astrophysical Journal Letters, 855, L15

\bibitem[{{Kremer} {et~al.}(2020){Kremer}, {Ye}, {Chatterjee}, {Rodriguez}, \&
  {Rasio}}]{kremer2020iau}
{Kremer}, K., {Ye}, C.~S., {Chatterjee}, S., {Rodriguez}, C.~L., \& {Rasio},
  F.~A. 2020, in Star Clusters: From the Milky Way to the Early Universe, ed.
  A.~{Bragaglia}, M.~{Davies}, A.~{Sills}, \& E.~{Vesperini}, Vol. 351,
  357--366

\bibitem[{Kremer {et~al.}(2020)Kremer, Claire, Rui, Weatherford, Chatterjee,
  Fragione, Rodriguez, Spera, \& Rasio}]{kremer2019cmcgrid}
Kremer, K., Claire, S.~Y., Rui, N.~Z., {et~al.} 2020, The Astrophysical Journal
  Supplement Series, 247, 48

\bibitem[{Kruijssen {et~al.}(2019)Kruijssen, Pfeffer, Reina-Campos, Crain, \&
  Bastian}]{kruijssen2019formation}
Kruijssen, J.~D., Pfeffer, J.~L., Reina-Campos, M., Crain, R.~A., \& Bastian,
  N. 2019, Monthly Notices of the Royal Astronomical Society, 486, 3180

\bibitem[{Lorimer(2008)}]{lorimer2008binary}
Lorimer, D.~R. 2008, Living Reviews in Relativity, 11, 8

\bibitem[{Lynch {et~al.}(2012)Lynch, Freire, Ransom, \&
  Jacoby}]{lynch2012timing}
Lynch, R.~S., Freire, P.~C., Ransom, S.~M., \& Jacoby, B.~A. 2012, The
  Astrophysical Journal, 745, 109

\bibitem[{Mackey \& Gilmore(2003{\natexlab{a}})}]{lmc2003surface}
Mackey, A., \& Gilmore, G. 2003{\natexlab{a}}, Monthly Notices of the Royal
  Astronomical Society, 338, 85

\bibitem[{Mackey \& Gilmore(2003{\natexlab{b}})}]{smc2003surface}
---. 2003{\natexlab{b}}, Monthly Notices of the Royal Astronomical Society,
  338, 120

\bibitem[{Mackey \& Gilmore(2003{\natexlab{c}})}]{fornax2003surface}
---. 2003{\natexlab{c}}, Monthly Notices of the Royal Astronomical Society,
  340, 175

\bibitem[{Mackey {et~al.}(2008)Mackey, Wilkinson, Davies, \&
  Gilmore}]{mackey2008black}
Mackey, A., Wilkinson, M., Davies, M.~B., \& Gilmore, G. 2008, Monthly Notices
  of the Royal Astronomical Society, 386, 65

\bibitem[{Maoz {et~al.}(2014)Maoz, Mannucci, \&
  Nelemans}]{maoz2014observational}
Maoz, D., Mannucci, F., \& Nelemans, G. 2014, Annual Review of Astronomy and
  Astrophysics, 52, 107

\bibitem[{McLaughlin \& van~der Marel(2005)}]{mclaughlin2005resolved}
McLaughlin, D.~E., \& van~der Marel, R.~P. 2005, The Astrophysical Journal
  Supplement Series, 161, 304

\bibitem[{Milone {et~al.}(2012)Milone, Piotto, Bedin, Aparicio, Anderson,
  Sarajedini, Marino, Moretti, Davies, Chaboyer, {et~al.}}]{milone2012acs}
Milone, A., Piotto, G., Bedin, L., {et~al.} 2012, Astronomy \& Astrophysics,
  540, A16

\bibitem[{Noyola \& Gebhardt(2006)}]{noyola2006surface}
Noyola, E., \& Gebhardt, K. 2006, The Astronomical Journal, 132, 447

\bibitem[{Oliphant(2006)}]{numpy}
Oliphant, T.~E. 2006

\bibitem[{Paresce {et~al.}(1995)Paresce, De~Marchi, \&
  Romaniello}]{paresce1995very}
Paresce, F., De~Marchi, G., \& Romaniello, M. 1995, The Astrophysical Journal,
  440, 216

\bibitem[{Pattabiraman {et~al.}(2013)Pattabiraman, Umbreit, Liao, Choudhary,
  Kalogera, Memik, \& Rasio}]{pattabiraman2013parallel}
Pattabiraman, B., Umbreit, S., Liao, W.-k., {et~al.} 2013, The Astrophysical
  Journal Supplement Series, 204, 15

\bibitem[{Perera {et~al.}(2017)Perera, Stappers, Lyne, Bassa, Cognard,
  Guillemot, Kramer, Theureau, \& Desvignes}]{perera2017evidence}
Perera, B., Stappers, B., Lyne, A., {et~al.} 2017, Monthly Notices of the Royal
  Astronomical Society, 468, 2114

\bibitem[{P{\'e}rez \& Granger(2007)}]{ipython}
P{\'e}rez, F., \& Granger, B.~E. 2007, Computing in Science \& Engineering, 9,
  21.
\newblock \url{https://aip.scitation.org/doi/abs/10.1109/MCSE.2007.53}

\bibitem[{Peuten {et~al.}(2014)Peuten, Brockamp, Kuepper, \&
  Kroupa}]{peuten2014puzzling}
Peuten, M., Brockamp, M., Kuepper, A.~H., \& Kroupa, P. 2014, The Astrophysical
  Journal, 795, 116

\bibitem[{Piatti(2018)}]{piatti2018extended}
Piatti, A.~E. 2018, Monthly Notices of the Royal Astronomical Society, 473, 492

\bibitem[{Pooley {et~al.}(2003)Pooley, Lewin, Anderson, Baumgardt, Filippenko,
  Gaensler, Homer, Hut, Kaspi, Makino, {et~al.}}]{pooley2003dynamical}
Pooley, D., Lewin, W.~H., Anderson, S.~F., {et~al.} 2003, The Astrophysical
  Journal Letters, 591, L131

\bibitem[{Portegies~Zwart {et~al.}(2010)Portegies~Zwart, McMillan, \&
  Gieles}]{portegies2010young}
Portegies~Zwart, S.~F., McMillan, S.~L., \& Gieles, M. 2010, Annual review of
  astronomy and astrophysics, 48, 431

\bibitem[{Robinson(1976)}]{robinson1976structure}
Robinson, E.~L. 1976, Annual review of astronomy and astrophysics, 14, 119

\bibitem[{Rodrigo \& Solano(2013)}]{rodrigo2013filter}
Rodrigo, C., \& Solano, E. 2013, The Filter Profile Service Access Protocol, ,

\bibitem[{Rodrigo {et~al.}(2017)Rodrigo, Solano, \& Bayo}]{rodrigo2017svo}
Rodrigo, C., Solano, E., \& Bayo, A. 2017, The SVO Filter Profile Service, ,

\bibitem[{Rodriguez {et~al.}(2018)Rodriguez, Amaro-Seoane, Chatterjee, \&
  Rasio}]{rodriguez2018post}
Rodriguez, C.~L., Amaro-Seoane, P., Chatterjee, S., \& Rasio, F.~A. 2018,
  Physical review letters, 120, 151101

\bibitem[{Rodriguez {et~al.}(2016)Rodriguez, Morscher, Wang, Chatterjee, Rasio,
  \& Spurzem}]{rodriguez2016million}
Rodriguez, C.~L., Morscher, M., Wang, L., {et~al.} 2016, Monthly Notices of the
  Royal Astronomical Society, 463, 2109

\bibitem[{Roh {et~al.}(2011)Roh, Lee, Joo, Han, Sohn, \& Lee}]{roh2011two}
Roh, D.-G., Lee, Y.-W., Joo, S.-J., {et~al.} 2011, The Astrophysical Journal
  Letters, 733, L45

\bibitem[{Rui {et~al.}(2021)Rui, Kremer, Weatherford, Chatterjee, Rasio,
  Rodriguez, \& Ye}]{codetag}
Rui, N.~Z., Kremer, K., Weatherford, N.~C., {et~al.} 2021, cmctoolkit,  Zenodo,
  doi:10.5281/zenodo.4579950

\bibitem[{Sarajedini \& Demarque(1990)}]{sarajedini1990new}
Sarajedini, A., \& Demarque, P. 1990, The Astrophysical Journal, 365, 219

\bibitem[{Servillat {et~al.}(2008)Servillat, Webb, \&
  Barret}]{servillat2008xmm}
Servillat, M., Webb, N., \& Barret, D. 2008, Astronomy \& Astrophysics, 480,
  397

\bibitem[{Sohn(2018)}]{sohn2018new}
Sohn, S.~T. 2018, stis, 2

\bibitem[{Sollima \& Baumgardt(2017)}]{sollima2017global}
Sollima, A., \& Baumgardt, H. 2017, Monthly Notices of the Royal Astronomical
  Society, 471, 3668

\bibitem[{Stodolkiewicz(1986)}]{stodolkiewicz1986dynamical}
Stodolkiewicz, J. 1986, Acta Astronomica, 36, 19

\bibitem[{Tam {et~al.}(2011)Tam, Kong, Hui, Cheng, Li, \& Lu}]{tam2011gamma}
Tam, P., Kong, A., Hui, C., {et~al.} 2011, The Astrophysical Journal, 729, 90

\bibitem[{Taylor {et~al.}(2001)Taylor, Grindlay, Edmonds, \&
  Cool}]{taylor2001helium}
Taylor, J., Grindlay, J., Edmonds, P., \& Cool, A. 2001, The Astrophysical
  Journal Letters, 553, L169

\bibitem[{{The Astropy Collaboration} {et~al.}(2013){The Astropy
  Collaboration}, {Robitaille, Thomas P.}, {Tollerud, Erik J.}, {Greenfield,
  Perry}, {Droettboom, Michael}, {Bray, Erik}, {Aldcroft, Tom}, {Davis, Matt},
  {Ginsburg, Adam}, {Price-Whelan, Adrian M.}, {Kerzendorf, Wolfgang E.},
  {Conley, Alexander}, {Crighton, Neil}, {Barbary, Kyle}, {Muna, Demitri},
  {Ferguson, Henry}, {Grollier, Frédéric}, {Parikh, Madhura M.}, {Nair,
  Prasanth H.}, {Günther, Hans M.}, {Deil, Christoph}, {Woillez, Julien},
  {Conseil, Simon}, {Kramer, Roban}, {Turner, James E. H.}, {Singer, Leo},
  {Fox, Ryan}, {Weaver, Benjamin A.}, {Zabalza, Victor}, {Edwards, Zachary I.},
  {Azalee Bostroem, K.}, {Burke, D. J.}, {Casey, Andrew R.}, {Crawford, Steven
  M.}, {Dencheva, Nadia}, {Ely, Justin}, {Jenness, Tim}, {Labrie, Kathleen},
  {Lim, Pey Lian}, {Pierfederici, Francesco}, {Pontzen, Andrew}, {Ptak, Andy},
  {Refsdal, Brian}, {Servillat, Mathieu}, \& {Streicher, Ole}}]{astropy_collab}
{The Astropy Collaboration}, {Robitaille, Thomas P.}, {Tollerud, Erik J.},
  {et~al.} 2013, A\&A, 558, A33.
\newblock \url{https://doi.org/10.1051/0004-6361/201322068}

\bibitem[{Trager {et~al.}(1995)Trager, King, \&
  Djorgovski}]{trager1995catalogue}
Trager, S., King, I.~R., \& Djorgovski, S. 1995, The Astronomical Journal, 109,
  218

\bibitem[{Umbreit {et~al.}(2012)Umbreit, Fregeau, Chatterjee, \&
  Rasio}]{umbreit2012monte}
Umbreit, S., Fregeau, J.~M., Chatterjee, S., \& Rasio, F.~A. 2012, The
  Astrophysical Journal, 750, 31

\bibitem[{Wang {et~al.}(2016)Wang, Spurzem, Aarseth, Giersz, Askar, Berczik,
  Naab, Schadow, \& Kouwenhoven}]{wang2016dragon}
Wang, L., Spurzem, R., Aarseth, S., {et~al.} 2016, Monthly Notices of the Royal
  Astronomical Society, 458, 1450

\bibitem[{Watkins {et~al.}(2015)Watkins, van~der Marel, Bellini, \&
  Anderson}]{watkins2015hubble}
Watkins, L.~L., van~der Marel, R.~P., Bellini, A., \& Anderson, J. 2015, The
  Astrophysical Journal, 803, 29

\bibitem[{Weatherford {et~al.}(2020)Weatherford, Chatterjee, Kremer, \&
  Rasio}]{weatherford2019dynamical}
Weatherford, N.~C., Chatterjee, S., Kremer, K., \& Rasio, F.~A. 2020, The
  Astrophysical Journal, 898, 162

\bibitem[{Weatherford {et~al.}(2018)Weatherford, Chatterjee, Rodriguez, \&
  Rasio}]{weatherford2018predicting}
Weatherford, N.~C., Chatterjee, S., Rodriguez, C.~L., \& Rasio, F.~A. 2018, The
  Astrophysical Journal, 864, 13

\bibitem[{Weatherford {et~al.}(2021)Weatherford, Fragione, Kremer, Chatterjee,
  Ye, Rodriguez, \& Rasio}]{weatherford2101black}
Weatherford, N.~C., Fragione, G., Kremer, K., {et~al.} 2021, arXiv preprint
  arXiv:2101.02217

\bibitem[{{W}es {M}c{K}inney(2010)}]{mckinney-proc-scipy-2010}
{W}es {M}c{K}inney. 2010, in {P}roceedings of the 9th {P}ython in {S}cience
  {C}onference, ed. {S}t\'efan van~der {W}alt \& {J}arrod {M}illman, 56 -- 61

\bibitem[{Ye {et~al.}(2019{\natexlab{a}})Ye, Kremer, Chatterjee, Rodriguez, \&
  Rasio}]{ye2019millisecond}
Ye, C.~S., Kremer, K., Chatterjee, S., Rodriguez, C.~L., \& Rasio, F.~A.
  2019{\natexlab{a}}, The Astrophysical Journal, 877, 122

\bibitem[{Ye {et~al.}(2019{\natexlab{b}})Ye, Kremer, Chatterjee, Rodriguez, \&
  Rasio}]{claire2019millisecond}
---. 2019{\natexlab{b}}, The Astrophysical Journal, 877, 122

\bibitem[{Zonoozi {et~al.}(2014)Zonoozi, Haghi, K{\"u}pper, Baumgardt, Frank,
  \& Kroupa}]{zonoozi2014direct}
Zonoozi, A.~H., Haghi, H., K{\"u}pper, A.~H., {et~al.} 2014, Monthly Notices of
  the Royal Astronomical Society, 440, 3172

\bibitem[{Zonoozi {et~al.}(2011)Zonoozi, K{\"u}pper, Baumgardt, Haghi, Kroupa,
  \& Hilker}]{zonoozi2011direct}
Zonoozi, A.~H., K{\"u}pper, A.~H., Baumgardt, H., {et~al.} 2011, Monthly
  Notices of the Royal Astronomical Society, 411, 1989

\end{thebibliography}

\appendix 

\section{Best Fits to $59$ Observed GCs}\label{appendix}

In this Appendix, we include a table of GCs with available SBPs and VDPs such that both have more than $5$ data points.
For GCs for which at least one snapshot with $s=\max\left(\tilde{\chi}_{\mathrm{SBP}}^2,\tilde{\chi}_{\mathrm{VDP}}^2\right)<10$, we report in Table \ref{big_table} all model parameters with well-fitting snapshots, together with the number $N_{\mathrm{good}}$ of well-fitting snapshots and the fitting and diagnostic parameters $s$, $\tilde{\chi}_{\mathrm{SBP}}^2$ and $\tilde{\chi}_{\mathrm{VDP}}^2$ themselves, $\tilde{\beta}_{\mathrm{SBP}}^2$, and $\tilde{\beta}_{\mathrm{VDP}}^2$.
For other GCs, only the model containing the best-fitting snapshot is shown together with the same quantities (notably, the ``best fit'' in these cases is not considered a ``good fit'').

\startlongtable
\begin{deluxetable*}{l||cc|cccc||c|ccc|cc} \label{big_table}
\centerwidetable
\tabletypesize{\scriptsize}
\tablecolumns{12}
\tablewidth{0pt}
 \tablecaption{Best-fitting model parameters, fitting figures of merit, and estimated black hole populations for Milky Way GCs
 \label{tab:clusterfits}}
 \tablehead{
 \colhead{Name} &
 \colhead{$N_{\mathrm{SBP}}$} &
 \colhead{$N_{\mathrm{VDP}}$} &
 \colhead{$r_v$} &
 \colhead{$r_g$} &
 \colhead{$[\mathrm{M}/\mathrm{H}]$} &
 \colhead{$N$} &
 \colhead{$N_{\mathrm{good}}$} &
 \colhead{$s$} &
 \colhead{$\tilde{\chi}_{\mathrm{SBP}}^2$} &
 \colhead{$\tilde{\chi}_{\mathrm{VDP}}^2$} &
 \colhead{$\tilde{\beta}_{\mathrm{SBP}}^2$} &
 \colhead{$\tilde{\beta}_{\mathrm{VDP}}^2$} \\
\colhead{} & \colhead{} & \colhead{} & \colhead{pc} & \colhead{kpc} & \colhead{} & \colhead{$\times10^5$} &
\colhead{} & \colhead{} & \colhead{} & \colhead{} & \colhead{} & \colhead{}
 }
\startdata
NGC 6553 & 105 & 9 & 2.0 & 2 & 0 & 16 & 73 & 0.92 & 0.92 & 0.91 & -0.13 & -0.70 \\
 &  &  &2.0 & 2 & 0 & 8 & 3 & 9.02 & 7.01 & 9.02 & -6.97 & 9.02 \\
NGC 4372 & 23 & 12 & 4.0 & 8 & -2 & 8 & 30 & 1.06 & 1.06 & 0.53 & 0.34 & -0.03 \\
 &  &  &2.0 & 8 & -2 & 8 & 30 & 5.34 & 5.34 & 0.95 & 5.34 & -0.63 \\
NGC 6352 & 56 & 6 & 2.0 & 2 & -1 & 4 & 11 & 1.56 & 1.56 & 0.57 & 1.06 & 0.57 \\
NGC 288 & 92 & 40 & 4.0 & 8 & -1 & 4 & 16 & 1.88 & 1.23 & 1.88 & 0.70 & 1.79 \\
NGC 6723 & 189 & 13 & 2.0 & 2 & -1 & 8 & 29 & 1.88 & 1.88 & 0.82 & -1.10 & 0.40 \\
NGC 6569 & 142 & 5 & 1.0 & 2 & -1 & 16 & 64 & 2.34 & 1.30 & 2.34 & 0.26 & -1.76 \\
 &  &  &1.0 & 2 & -1 & 8 & 30 & 3.62 & 1.90 & 3.62 & -0.63 & 3.62 \\
NGC 6656 & 146 & 53 & 1.0 & 2 & -2 & 16 & 70 & 3.17 & 1.74 & 3.17 & 1.00 & 2.78 \\
NGC 5897 & 90 & 7 & 4.0 & 8 & -2 & 4 & 14 & 3.84 & 1.46 & 3.84 & 0.02 & 3.84 \\
NGC 6779 & 152 & 6 & 1.0 & 8 & -2 & 8 & 16 & 3.68 & 3.68 & 2.28 & 3.16 & -1.91 \\
 &  &  &2.0 & 8 & -2 & 8 & 14 & 4.04 & 4.04 & 1.26 & -0.06 & -0.79 \\
 &  &  &2.0 & 8 & -2 & 4 & 7 & 4.70 & 2.32 & 4.70 & -1.68 & 4.70 \\
NGC 1904 & 355 & 11 & 1.0 & 20 & -2 & 8 & 7 & 4.21 & 4.21 & 2.92 & 2.68 & -2.92 \\
NGC 5986 & 85 & 8 & 1.0 & 2 & -2 & 16 & 20 & 4.36 & 4.36 & 0.91 & -0.94 & 0.13 \\
NGC 6681 & 150 & 40 & 0.5 & 2 & -2 & 8 & 49 & 4.37 & 3.71 & 4.37 & 3.60 & 4.27 \\
NGC 6541 & 128 & 13 & 2.0 & 2 & -2 & 16 & 36 & 4.38 & 3.11 & 4.38 & -1.18 & 1.69 \\
 &  &  &1.0 & 2 & -2 & 16 & 76 & 6.91 & 1.79 & 6.91 & 0.75 & -1.91 \\
NGC 5024 & 223 & 8 & 2.0 & 20 & -2 & 16 & 75 & 5.02 & 5.02 & 2.14 & -3.41 & -1.19 \\
 &  &  &1.0 & 20 & -2 & 16 & 3 & 9.23 & 9.23 & 2.46 & 0.76 & -1.50 \\
NGC 6293 & 234 & 7 & 1.0 & 2 & -2 & 8 & 8 & 5.06 & 5.06 & 1.35 & -1.21 & 1.04 \\
 &  &  &0.5 & 2 & -2 & 8 & 51 & 5.26 & 5.26 & 1.92 & 2.29 & 1.89 \\
NGC 6712 & 58 & 8 & 1.0 & 2 & -1 & 8 & 30 & 5.97 & 5.97 & 3.79 & -0.58 & -3.79 \\
 &  &  &2.0 & 2 & -1 & 8 & 17 & 6.06 & 6.06 & 0.39 & -5.26 & -0.36 \\
NGC 6397 & 344 & 50 & 1.0 & 8 & -2 & 4 & 11 & 6.16 & 6.16 & 4.78 & 4.24 & 3.87 \\
NGC 3201 & 84 & 32 & 2.0 & 8 & -2 & 4 & 8 & 6.17 & 1.51 & 6.17 & 0.90 & 5.99 \\
 &  &  &2.0 & 8 & -2 & 8 & 7 & 8.64 & 8.24 & 8.64 & 7.11 & -8.60 \\
NGC 6539 & 94 & 5 & 1.0 & 2 & -1 & 16 & 64 & 3.99 & 3.38 & 3.99 & -2.41 & -3.99 \\
 &  &  &1.0 & 2 & -1 & 8 & 13 & 6.32 & 6.32 & 0.59 & -3.49 & 0.59 \\
NGC 6121 & 230 & 26 & 1.0 & 8 & -1 & 4 & 16 & 6.86 & 4.09 & 6.86 & 2.63 & 5.85 \\
NGC 1261 & 129 & 10 & 2.0 & 20 & -1 & 8 & 12 & 7.05 & 7.05 & 1.10 & -0.37 & -0.86 \\
 &  &  &1.0 & 20 & -1 & 8 & 1 & 8.89 & 8.89 & 2.12 & 7.48 & -1.71 \\
 &  &  &2.0 & 20 & -1 & 4 & 1 & 9.70 & 9.70 & 5.12 & -9.57 & 5.12 \\
NGC 1851 & 102 & 42 & 0.5 & 20 & -1 & 16 & 8 & 7.66 & 2.08 & 7.66 & -1.57 & -6.31 \\
NGC 6496 & 30 & 9 & 4.0 & 2 & 0 & 8 & 12 & 7.83 & 5.65 & 7.83 & 5.65 & -5.00 \\
Ter 5 & 62 & 6 & 1.0 & 2 & 0 & 16 & 44 & 8.43 & 2.65 & 8.43 & 0.91 & 8.43 \\
NGC 5286 & 98 & 10 & 1.0 & 8 & -2 & 16 & 3 & 8.91 & 8.91 & 0.49 & -0.76 & 0.29 \\
 &  &  &1.0 & 8 & -2 & 8 & 10 & 9.93 & 2.23 & 9.93 & -1.04 & 9.92 \\
NGC 6171 & 102 & 18 & 2.0 & 2 & -1 & 8 & 1 & 9.32 & 9.32 & 4.44 & -2.90 & -4.44 \\
\hline
NGC 6304 & 108 & 6 & 2.0 & 2 & 0 & 8 & 0 & 10.68 & 10.68 & 1.40 & -4.16 & -1.32 \\
NGC 6205 & 128 & 14 & 1.0 & 8 & -2 & 16 & 0 & 11.43 & 11.43 & 1.95 & 10.39 & 1.28 \\
NGC 6809 & 115 & 13 & 2.0 & 2 & -2 & 8 & 0 & 11.77 & 11.77 & 2.69 & 9.15 & 0.83 \\
NGC 6624 & 279 & 33 & 0.5 & 2 & 0 & 8 & 0 & 11.95 & 11.95 & 1.71 & -2.65 & -0.28 \\
NGC 6218 & 144 & 12 & 2.0 & 2 & -1 & 8 & 0 & 12.07 & 12.07 & 5.05 & -6.16 & -5.05 \\
NGC 362 & 241 & 53 & 0.5 & 8 & -1 & 16 & 0 & 12.15 & 1.94 & 12.15 & -0.12 & -10.79 \\
NGC 5272 & 91 & 21 & 1.0 & 8 & -2 & 16 & 0 & 12.42 & 12.42 & 1.68 & 10.71 & -0.52 \\
NGC 6366 & 28 & 9 & 2.0 & 2 & -1 & 4 & 0 & 13.75 & 13.75 & 1.24 & 12.54 & -1.18 \\
NGC 4590 & 240 & 7 & 2.0 & 8 & -2 & 8 & 0 & 13.96 & 13.96 & 8.47 & 12.98 & -8.47 \\
NGC 6626 & 326 & 11 & 0.5 & 2 & -1 & 16 & 0 & 14.27 & 14.27 & 0.90 & 14.08 & -0.42 \\
NGC 6402 & 84 & 11 & 1.0 & 2 & -1 & 16 & 0 & 14.31 & 14.31 & 12.31 & -13.00 & 12.31 \\
NGC 6362 & 58 & 31 & 2.0 & 8 & -1 & 4 & 0 & 14.74 & 14.74 & 1.34 & 9.28 & 0.48 \\
NGC 7089 & 269 & 21 & 1.0 & 8 & -2 & 16 & 0 & 15.41 & 8.84 & 15.41 & -6.97 & 15.41 \\
NGC 6273 & 125 & 9 & 1.0 & 2 & -2 & 16 & 0 & 15.66 & 9.76 & 15.66 & -8.64 & 15.66 \\
NGC 6522 & 274 & 11 & 0.5 & 2 & -1 & 16 & 0 & 18.60 & 18.60 & 1.75 & 16.70 & -1.68 \\
NGC 7099 & 297 & 22 & 1.0 & 8 & -2 & 4 & 0 & 18.68 & 9.98 & 18.68 & 5.20 & 18.53 \\
NGC 6752 & 334 & 50 & 1.0 & 8 & -2 & 8 & 0 & 19.90 & 5.12 & 19.90 & -0.16 & 19.72 \\
NGC 5904 & 125 & 52 & 1.0 & 8 & -1 & 16 & 0 & 23.14 & 23.14 & 6.81 & -21.65 & -6.45 \\
NGC 5927 & 57 & 40 & 2.0 & 2 & 0 & 8 & 0 & 24.27 & 20.75 & 24.27 & -16.31 & 24.25 \\
NGC 6093 & 267 & 11 & 1.0 & 2 & -2 & 16 & 0 & 26.16 & 26.16 & 11.22 & 6.34 & 11.19 \\
NGC 7078 & 405 & 50 & 1.0 & 8 & -2 & 16 & 0 & 29.27 & 20.50 & 29.27 & -18.84 & 28.64 \\
NGC 6535 & 58 & 8 & 4.0 & 2 & -2 & 16 & 0 & 30.08 & 15.36 & 30.08 & 15.18 & -30.08 \\
NGC 5824 & 81 & 5 & 1.0 & 20 & -2 & 32 & 0 & 32.22 & 32.22 & 0.53 & 0.46 & 0.53 \\
NGC 6254 & 161 & 23 & 1.0 & 2 & -2 & 8 & 0 & 33.53 & 33.53 & 2.18 & 26.58 & 0.24 \\
NGC 6341 & 99 & 39 & 1.0 & 8 & -2 & 16 & 0 & 58.86 & 58.86 & 5.23 & -57.97 & -5.19 \\
NGC 6715 & 227 & 35 & 0.5 & 20 & -1 & 16 & 0 & 115.00 & 115.00 & 78.23 & -115.00 & 78.23 \\
NGC 2419 & 139 & 5 & 2.0 & 20 & -2 & 32 & 0 & 119.26 & 119.26 & 3.05 & 26.26 & -3.05 \\
NGC 2808 & 304 & 48 & 0.5 & 8 & -1 & 16 & 0 & 167.77 & 30.86 & 167.77 & -30.15 & 167.77 \\
NGC 6266 & 227 & 42 & 0.5 & 2 & -1 & 16 & 0 & 199.06 & 17.86 & 199.06 & -17.84 & 199.06 \\
NGC 6388 & 193 & 42 & 0.5 & 2 & -1 & 16 & 0 & 206.15 & 37.78 & 206.15 & -36.32 & 206.15 \\
NGC 104 & 204 & 62 & 0.5 & 8 & -1 & 16 & 0 & 222.64 & 222.64 & 91.41 & -222.64 & 91.41 \\
NGC 6441 & 158 & 37 & 1.0 & 2 & 0 & 16 & 0 & 238.98 & 187.23 & 238.98 & -187.23 & 238.98 \\
NGC 5139 & 73 & 65 & 1.0 & 8 & -2 & 16 & 0 & 1086.16 & 151.20 & 1086.16 & -151.20 & 1086.16 \\
\hline
\enddata
\tablecomments{For $59$ Milky Way GCs, we report the number $N_{\mathrm{SBP}}$ of data points in the SBP, number $N_{\mathrm{VDP}}$ of data points in the VDP, initial virial radius, galactocentric distance, metallicity, and initial particle number of the well-fitting or best-fitting model(s) (depending on whether the GC is well-fit), the number $N_{\mathrm{good}}$ of well-fitting snapshots, $s$, $\tilde{\chi}_{\mathrm{SBP}}^2$, $\tilde{\chi}_{\mathrm{VDP}}^2$, $\tilde{\beta}_{\mathrm{SBP}}^2$, and $\tilde{\beta}_{\mathrm{VDP}}^2$.}
\end{deluxetable*}

$ $ % This is super dumb but it works for stopping the table from getting cut off

\end{document}